\documentclass[sigconf]{acmart}
\usepackage{booktabs} 
\usepackage{float}
\usepackage{multirow}
\usepackage{graphicx}
\usepackage[utf8]{inputenc}
\usepackage{array}
\usepackage{soul}
\usepackage{graphicx}
\usepackage{xfrac}

\usepackage{draftwatermark}
\SetWatermarkText{Accepted Manuscript}
\SetWatermarkScale{.25}
\SetWatermarkAngle{0}

\newcolumntype{P}[1]{>{\centering\arraybackslash}p{#1}}
\newcolumntype{M}[1]{>{\centering\arraybackslash}m{#1}}

\AtBeginDocument{%
  \providecommand\BibTeX{{%
    \normalfont B\kern-0.5em{\scshape i\kern-0.25em b}\kern-0.8em\TeX}}}

\setcopyright{acmcopyright}
\copyrightyear{2021}
\acmYear{2021}
\setcopyright{acmlicensed}\acmConference[CHI '21]{CHI Conference on Human Factors in Computing Systems}{May 8--13, 2021}{Yokohama, Japan}
\acmBooktitle{CHI Conference on Human Factors in Computing Systems (CHI '21), May 8--13, 2021, Yokohama, Japan}
\acmPrice{15.00}
\acmDOI{10.1145/3411764.3445397}
\acmISBN{978-1-4503-8096-6/21/05}

\begin{document}

\title[How Do YouTubers Help with COVID-19 Loneliness?]{\#StayHome \#WithMe: How Do YouTubers Help with COVID-19 Loneliness?}

\author{Shuo Niu}
\orcid{https://orcid.org/0000-0002-8316-4785}
\author{Ava Bartolome}
\authornote{Student authors contributed equally to this research.}
\author{Cat Mai}
\authornotemark[1]
\author{Nguyen B. Ha}
\authornotemark[1]
\email{{shniu, abartolome, cmai, joha}@clarku.edu}
\affiliation{%
  \institution{Clark University}
  \streetaddress{950 Main St.}
  \city{Worcester}
  \state{MA}
  \country{USA}
  \postcode{01610}
}
\renewcommand{\shortauthors}{Niu, Bartolome, Mai, and Ha}

\begin{abstract}
    Loneliness threatens public mental wellbeing during COVID-19. In response, YouTube creators participated in the \#StayHome \#WithMe movement (SHWM) and made myriad videos for people experiencing loneliness or boredom at home. User-shared videos generate parasocial attachment and virtual connectedness. However, there is limited knowledge of how creators contributed videos during disasters to provide social provisions as disaster-relief. Grounded on Weiss's loneliness theory, this work analyzed 1488 SHWM videos to examine video sharing as a pathway to social provisions. Findings suggested that skill and knowledge sharing, entertaining arts, homelife activities, live chatting, and gameplay were the most popular video styles. YouTubers utilized parasocial relationships to form a space for staying away from the disaster. SHWM YouTubers provided friend-like, mentor-like, and family-like provisions through videos in different styles. Family-like provisions led to the highest overall viewer engagement. Based on the findings, design implications for supporting viewers' mental wellbeing in disasters are discussed.
\end{abstract}

\ccsdesc[500]{Human-centered computing~Empirical studies in collaborative and social computing}

\keywords{YouTube, video sharing, parasocial, social provisions, disaster, loneliness}

\maketitle

\section{Introduction}
In 2020, the COVID-19 pandemic swept the globe and forced billions of people to stay home for months. Social media became an alternative venue for people to stay connected and cope with loneliness during this global crisis. Loneliness is a significant public mental\-health issue during COVID-19 \cite{PalgiLonelinesspandemic, GroarkeLonelinessPandemic}. YouTube, as the largest video sharing platform, called on YouTube creators (YouTubers) to join the \#StayHome \#WithMe movement (SHWM) by creating ``\textit{content that will entertain, inform, and connect with people who are social distancing during this pandemic}''\footnote{https://services.google.com/fh/files/misc/stay\_home\_with\_me.pdf}. This hashtag movement attracted a lot of video artists and creators, which witnessed a 600\% increase in viewership\footnote{https://www.theverge.com/2020/3/27/21197642/youtube-with-me-style-videos-\\views-coronavirus-cook-workout-study-home-beauty}. There is increasing attention in the literature on using social media to support public safety and wellbeing during disasters \cite{HoustonSocialMediaDisaster}. In HCI, researchers have explored the affordances of Twitter \cite{PalenTwitterDiaster}, Facebook \cite{BirdFloodingFacebook, ChauhanWildfireFacebook}, and Reddit \cite{YunanRedditCrisis} in supporting disaster awareness and communication. However, there is limited understanding of how YouTubers and video-sharing platforms can uniquely support public mental health during a long-term crisis. \#StayHome \#WithMe is a YouTubers' voluntary response to help people who might feel lonely or bored during COVID-19. This trending movement provides a valuable and unique lens to examine the social connections that YouTubers offered to mitigate loneliness during social distancing.
\par
Video-sharing platforms such as YouTube, Twitch, TikTok, and Facebook Videos saw a rapid increase in popularity in the past decade. In 2019, YouTube was the second most popular social media, and 73\% of adults have visited this video sharing platform \cite{perrin_anderson_2019}. The participatory culture on YouTube encourages grassroots to participate in video creation. In contrast to platforms like Twitter and Facebook, social interactions on YouTube are based on video itself rather than personal profiles and friending \cite{Burgess2018YouTube:Culture}. YouTubers self-construct their value by contributing unique videos and actively interacting with fans \cite{HuVideoSharingCommunity}. The regular postings of YouTubers make the audience form a one-sided sense of closeness to the YouTubers. Prior studies described the audience's social and emotional attachment to a video persona as ``parasocial relationship'' \cite{WohnParasocialInteraction}. The interactions with videos to generate parasocial relationship are called ``parasocial interactions'' \cite{HartmannParasocialInteractionWellbeing}. When other social interactions are diminished during social distancing, SHWM can be seen as an effort to supplement social connections through parasocial relationships. However, there is limited understanding of YouTubers' roles and YouTube's affordances in offering social connections and mitigating disaster loneliness. Considering YouTube's massive creator base and high popularity among young people \cite{perrin_anderson_2019}, the HCI community needs to address this knowledge gap to better design applications and services to support disaster mental health. This work explores this new social media phenomenon by collecting and analyzing SHWM video data. Grounded on Weiss's loneliness theory, this work examines what loneliness-supporting videos were created, how they sought to offer social provisions, and whether different social provisions affected viewer engagement. An overview of SHWM can provide new perspectives on YouTube's roles in disasters and inform social media platforms' socio-technical design for tackling loneliness. 
\par
8023 SHWM videos published between March 11th and May 15th were crawled from YouTube Data API, among which 1642 made by creators in the United States were annotated by Amazon Mechanical Turk (MTurk) participants. These videos came from 695 video makers and attracted 206,319,247 views. The analysis of SHWM videos was guided by Weiss's framework of social-emotional loneliness \cite{RussellWeissLoneliness}. Weiss's theory conceptualizes six core types of social provisions that people need to refrain loneliness (Table \ref{tab:weiss_theory}) -- \textit{attachment}, \textit{social integration}, \textit{reassurance of worth}, \textit{reliable alliance}, \textit{guidance}, and \textit{opportunity for nurturance}. Weiss' model incorporates the major elements of social relationships people need from families, friends, and mentors \cite{Cutrona1987TheStress}, which are also the connections that are likely to be absent during social distancing. YouTubers approach viewers by sharing original content and arousing viewers to parasocially interact. Viewers might generate feelings of connection and intimacy to the creators \cite{Marwick2015YouMedia, RasmussenParasocialInteractionYouTubeCelebrities} and interact with YouTubers through liking and commenting. Examining the styles of SHWM videos and analyzing what social provisions the videos sought to offer will deepen the understanding of YouTube's social functionalities in disasters and the categories of virtual connectedness provided by YouTubers. Grounded on Weiss's framework, this paper addresses three key research questions:

\begin{itemize}
    \item RQ1. What \#StayHome \#WithMe videos did YouTubers make and how they relate to the COVID-19 pandemic?
    \item RQ2. How did \#StayHome \#WithMe videos in different styles and mentioning COVID-19 in different degrees affect the social provisions in the video?
    \item RQ3. How did videos with different social provisions affect viewer engagement?
\end{itemize}
RQ1 is an initial exploration of SHWM video styles and if YouTubers chose to communicate the ongoing COVID-19 pandemic. Rooted in Weiss's loneliness theory, RQ2 explores whether and how the six social provisions were associated with videos in different styles and different COVID-19 mentioning. Following the understanding of video styles and social provisions, RQ3 further explores whether offering various social provisions affected video popularity, viewers' activeness, and viewers' emotional expression in the comments. These metrics reflect the parasocial interactions with YouTubers. The authors found that SHWM videos primarily themed in: how-to videos of sharing skills and knowledge; entertaining content of music, arts, and performance; videos of homelife activities; chatting with the audience; and videos of gameplay. In contrast to other social media platforms where disaster information is intensively communicated and spread, these videos formed an online space where the disaster is not actively mentioned. The analysis of six social provisions in SHWM explained how parasocial relationships supplement social connections to mitigate loneliness. Most SHWM videos offered social integration by sharing interests and recreational activities. A large number of how-to videos primarily supported the guidance provision. Videos of homelife and chatting supported the provisions of attachment, nurturance, and alliance. Providing family-like social provisions had better overall viewer engagement with the video, despite their smaller proportions in SHWM. This work's findings foster new possibilities to design video-sharing-based applications and technologies to address disaster-related loneliness. 

\section{Related Work}

\subsection{Social Media and Disasters}
Much research has explored the use of different social media during disasters. Houston et al. reviewed prior works and concluded key affordances of social media include signaling and detecting disasters, sending and receiving requests for help, informing conditions, providing mental/behavioral support, etc. \cite{HoustonSocialMediaDisaster}. Lindsay summarized that social media such as Twitter and Facebook are primarily used to disseminate and receive disaster information and serve as an emergency management tool \cite{LindsaySocialMediaDisasters}. Studies examined the unique affordances of different social media in emergencies and disasters. For example, Twitter is a platform for sharing short and immediate disaster messages to raise public awareness of the critical situation \cite{PalenTwitterDiaster, NaamanTwitterSocialAwareness}. Facebook users seek official information and organize disaster-supporting communities \cite{BirdFloodingFacebook, ChauhanWildfireFacebook, SilverFacebookDisaster, PalenDisasterCommunity}. Reddit users perceive and speculate risks in the long run, or different regions \cite{YunanRedditCrisis}. YouTube videos educate the public as well as spread misinformation \cite{OwensDisasterYouTube, BaschYouTubeEbola}. Disasters are stressful events and can cause mental health problems for both people who are directly affected and the community at large \cite{RoudiniDisasterMentalHealthPreparedness}. Besides spreading disaster information, social media is also an outlet for expressing emotions \cite{KarmegamReviewMentalHealthSocialMedia, ProcopioCrisisSupportGroup} and receiving mental health service \cite{HoustonSocialMediaDisaster}. Therefore studies have looked into technologies to measure public mental health by mining social media data \cite{KarmegamReviewMentalHealthSocialMedia, ChoudhuryTwitterMentalHealth, ChoudhuryRedditMentalHealth}. Recent studies suggested that COVID social distancing caused loneliness and other mental health challenges to many who stayed at home \cite{LeiaCOVIDMentalHealth, PalgiLonelinesspandemic}, especially the young adults \cite{GroarkeLonelinessPandemic, EllisSocialStress}. However, the use of video-sharing for supporting mental health during disasters is under-explored \cite{perrin_anderson_2019}. There is limited understanding of how video creators contribute to a social media movement like \#StayHome \#WithMe to support disaster mental health and what affordances of the video-sharing platform and community can reduce disaster loneliness.

\subsection{YouTube and Parasocial Relationship}
In HCI and CSCW, understanding the unique characteristics and communication affordances of different platforms centers the research on social media \cite{PreeceOnlineCommunities, PalenInformatingCrisis}. After years of growth, video-sharing platforms like YouTube have developed their characteristics of platform cultures. Dijck noted that on YouTube, user-generated videos boost online production and distribute diverse content \cite{VanDijck2013TheMedia}. In contrast to platforms based on friending and networking, social interactions on YouTube rely on the video itself rather than offline relationships \cite{Burgess2018YouTube:Culture}. YouTubers interact with viewers through sharing new videos to cultivate relationships with fans \cite{HuVideoSharingCommunity, Hou2018SocialYouTube}. Horton and Wohl defined the one-sided intimacy generated by the ``conversational give-and-take'' with performers as \textit{parasocial relationship} \cite{HortonParasocialInteraction}. Video interactions that lead to parasocial relationships are called \textit{parasocial interactions} \cite{HortonParasocialInteraction, HartmannParasocialInteractionWellbeing}. Gardner found that people turn to parasocial relationships with a media figure when they need to regulate the needs of social belongingness \cite{GardnerSocialSnacking}. Hartmann concluded that parasocial relationships could provide social support and shield against the effects of exclusion and loneliness \cite{HartmannParasocialInteractionWellbeing}. Rotman et al. suggested YouTube users mostly focused on the parasocial interactions with the creator, rather than building a friend and community network like other social media \cite{PreeceYouTubeCommunity}. Wohn examined live streamers and found parasocial relationships correlate with viewers' emotional, instrumental, and financial support for the performers \cite{WohnParasocialInteraction}. Anjani et al. studied YouTube food-eating videos and suggested that they generate a sense of connectedness \cite{TangMukbang}. The parasocial interactions on YouTube lie in that YouTubers make original content to engage viewers, and viewers respond to and endorse YouTubers by video-watching, liking, and commenting \cite{Burgess2018YouTube:Culture, HuVideoSharingCommunity, Hou2018SocialYouTube, KhanSocialMediaEngagement}. \#StayHome \#WithMe is a movement in which YouTubers respond to the pervasive loneliness and use parasocial relationships to offer loneliness support. However, little is known about how parasocial relationships are embodied during an ongoing pandemic and their affordances to supplement social interactions. This work seeks to give a deeper understanding of the phenomenon of video-sharing for mitigating disaster loneliness. Considering video sharing and video interaction center the social activities on YouTube, it is essential to examine what SHWM videos did YouTubers provide, how different videos offer parasocial relationships to supplement social connections, and whether different social connections affect viewer interactions with the video. 

\subsection{Social Provisions of \#StayHome \#WithMe}
Loneliness is prevalent among people who experienced disasters and crises \cite{HugeliusLonelinessNaturalDisaster, MakSARSMentalHealth, PalgiLonelinesspandemic}. During COVID-19, technologies play a vital role in offering social support and helping people deal with loneliness and isolation \cite{LeiaCOVIDMentalHealth}. This work utilizes Weiss's theory of loneliness \cite{WeissEmotionalIsolation} to examine the role of SHWM videos in offering social provisions during COVID-19. Weiss conceptualized social and emotional loneliness and argued that people need six \textit{social provisions} to deal with loneliness \cite{RussellWeissLoneliness}; see Table \ref{tab:weiss_theory} for definitions, which are elaborated in section \ref{sub:social_provisions}. Cutrona and Russell examined Weiss's loneliness theory in psychological practice and identified the sources of social provisions\cite{Cutrona1987TheStress}. Attachment, reliable alliance, and opportunity for nurturance demonstrate intimacy or trust and are usually provided by family members \cite{Cutrona1987TheStress}. Social integration is usually obtained from friend relationships \cite{Cutrona1987TheStress}. Guidance is obtained from teachers, mentors, or parent figures \cite{Cutrona1987TheStress}. Reassurance of worth is a type of self-efficacy and self-esteem obtained by helping others and receiving acknowledgment \cite{Cutrona1987TheStress}. 

\par
\begin{table}[ht]
  \Description[Weiss's Framework of Social Provisions]{The definition of Weiss's Framework of Loneliness. The framework includes attachment, social integration, reassurance of worth, reliable alliance, guidance, and opportunity for nurturance.}
  \caption{Weiss's Framework of Social Provisions for Lone\-liness}
  \label{tab:weiss_theory}
  \scalebox{0.8}{
  \begin{tabular}{p{.26\linewidth}p{.6\linewidth}P{.2\linewidth}}
        \toprule
        Concept & Definition & Source\\
        \midrule
        attachment & A relationship in which people receives a sense of safety and security & family\\[4ex]
        
        social\linebreak integration & A network of relationships in which individuals share interests, concerns, and recreational activities & friend\\[6ex]
        
        reassurance\linebreak of worth & A relationships in which the person's skills and abilities are acknowledged & self \\[4ex]
        
        reliable alliance & A relationship in which one can count on assistance under any circumstances & family \\[4ex]
        
        guidance & A relationship with trustworthy and authoritative individuals who can provide advice and assistance & mentor\\[6ex]
        
        opportunity\linebreak for nurturance & A relationship in which the person feels responsible for the wellbeing of another & family\\
      \bottomrule
    \end{tabular}
    }
\end{table}
Weiss's loneliness framework has been used to study the social connections people can obtain from social media. The six social provisions were examined in the loneliness-support functions of Twitter \cite{GlasgowSocialMediaTraumaticEvents} and Facebook \cite{VitakFacebookBounding}. It was also used as a framework to study how parasocial relationships affect loneliness \cite{WangParasocialWeiss, HuCelebrityParasocialWeiss}. Weiss's framework described the necessary relationships people need in the context of loneliness \cite{Cutrona1987TheStress}, which are also the social supports that are likely to be absent during social distancing. YouTubers perform various kinds of affective relationships and serve as micro-celebrities to offer accessibility, authenticity, and connectedness to the audience \cite{RaunTransgenderYouTube, Marwick2015YouMedia}. Viewers experience YouTubers' characters as close friends or online family members to fulfill the need for social interaction \cite{RasmussenParasocialInteractionYouTubeCelebrities, KhanSocialMediaEngagement}. The parasocial relationships can be a source of social connections for people who need to improve emotional, cognitive, and behavioral health \cite{ReddySocialMediaDepression}. By examining Weiss's six provisions in SHWM videos, this work offers an overview of how YouTubers construct the parasocial relationships to supplement social provisions during social distancing. This understanding is essential to guide future research on viewers' perception of social relationships with YouTubers and the psychological and affective affordances of YouTube in reducing disaster loneliness.
\par
During COVID-19, viewers may choose to watch SHWM videos to obtain insufficient social provisions in real life to avoid loneliness \cite{GileParasocialInteraction}. This work uses engagement metrics to measure the viewers' participation in parasocial interactions. Prior studies suggested that participating in parasocial interactions can promote social connectedness and mitigate loneliness \cite{HartmannParasocialInteractionWellbeing, GardnerSocialSnacking}. Khan noted that viewers who want to socialize on YouTube were more likely to like/dislike and comment on the videos \cite{KhanSocialMediaEngagement}. Rasmussen found that viewer interactions such as commenting and sending messages to the YouTubers can simulate realistic social interactions \cite{RasmussenParasocialInteractionYouTubeCelebrities}. Wright et al. noticed that video-based platforms outperformed message-based media in promoting mental wellbeing \cite{WrightSocialMediaUse}. Similarly, on other social media, commenting and liking others' social media posts are positive indicators of mental wellbeing \cite{ParkFacebook, Verduyn2015PassiveEvidence.}. Studies showed that discussing common interests on social media helped users cope with loneliness \cite{FratamicoRedditLonely} and promoted comfort feelings \cite{OLearyChatMentalHealth}. This analysis provides a preliminary understanding of how six social provisions affect engagement in parasocial interactions. 

\section{Study Design}
\begin{figure}[ht]
    \Description[The analysis framework of the three research questions.]{RQ1 explores video styles and COVID-19 mentioning. RQ2 examines if how video styles and COVID mentioning affect social provisions. RQ3 examines how social provisions affect viewer engagement.}
    \includegraphics[width=.9\linewidth]{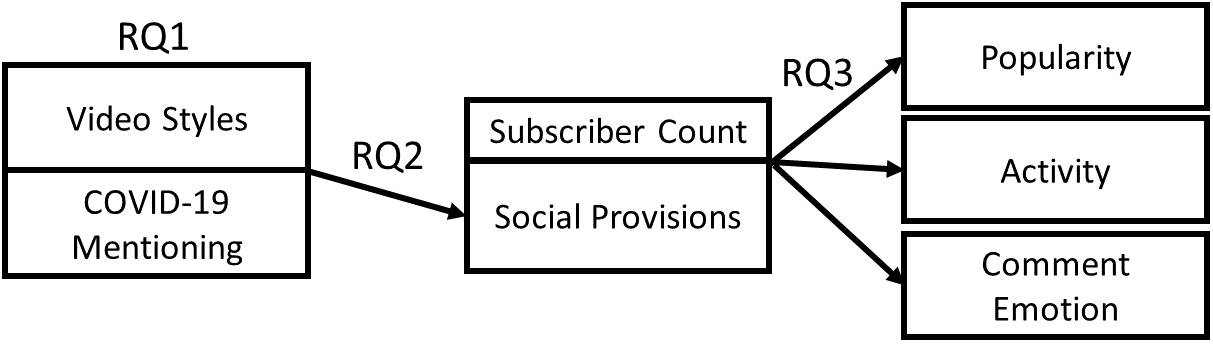}
    \caption{The analysis framework of the three research questions}
    \label{fig:concept}
\end{figure}

Grounded in prior research, this work uses SHWM as a lens to examine the social affordances of YouTube and YouTuber communities in supporting social connections and mitigating disaster loneliness. Based on Weiss's framework, the three research questions examine how SHWM videos provide social provisions and whether they affect viewer engagement. This section describes the structure of the study and the factors examined in the data analysis (Figure \ref{fig:concept}).
\vspace{-0.5cm}
\subsection{Video Styles and COVID-19 Mentioning}
Research on social media interaction needs to capture user behaviors and participation styles \cite{KapoorSocialMediaResearchPastFuture}. Addressing RQ1 provides an overview of how SHWM YouTubers crafted parasocial relationships for viewers who need to avoid or mitigate loneliness. Considering the diversity of \#StayHome \#WithMe participation, RQ1 examines SHWM videos by categorizing video styles and rating the degrees of COVID-19 mentioning. Video styles are the major themes and production styles of SHWM videos, including the video's topics and the main activities YouTubers performed. The categorization of SHWM video styles consisted of two steps. YouTube allows creators to specify their video content using the default YouTube categories, including ``music'', ``entertainment'', ``gaming'', etc. As an initial step, the category distribution was compared between SHWM videos and the overall YouTube category distribution in 2015 \cite{Che2015ACharacteristics}. This step provided a general understanding of what video categories trended in SHWM. In the second step, considering that the YouTube default categories lack coverage of topics such as homelife or chatting about the pandemic, grounded-theory methods were performed to derive video styles. The rating of COVID-19 mentioning pertained to the pandemic context of SHWM videos. Examining the levels to which COVID-19 was mentioned can unveil if YouTubers shared disaster-related information like other social media \cite{PalenTwitterDiaster, SilverFacebookDisaster, HoustonSocialMediaDisaster, LindsaySocialMediaDisasters}. These two factors allow the authors to infer how parasocial relationships were delivered during COVID-19 and whether they were affected by the disaster.

\subsection{Social Provisions}
\label{sub:social_provisions}
Following the categorization of SHWM videos, RQ2 measures whe\-ther and how SHWM videos provided different social provisions. This work adopts Weiss's loneliness theory as the theoretical framework. Weiss's theory depicts the categories of roles YouTubers can play via parasocial relationships in providing needed social provisions during social distancing. In Weiss's theory, \textit{attachment} is a social provision for a sense of safety and security. In SHWM, creators may give the viewers a sense of emotional security and closeness by showing intimate content such as activities at home or a face-to-face chat. \textit{Social integration} is a provision to share interests and concerns. By sharing videos of hobbies and interests, YouTubers can entertain the viewers and provide social integration. \textit{Reassurance of worth} emphasizes one's skills and abilities are acknowledged. YouTube has a culture of developing video creation skills and growing the subscriber community \cite{Burgess2018YouTube:Culture}. By sharing SHWM videos, YouTubers demonstrate their unique talents and be acknowledged by the viewers and gain reassurance. \textit{Reliable alliance} is a provision which the person can count on under any circumstances. To help people who are staying at home, YouTubers may express a willingness to help at any time. \textit{Guidance} is a provision through which people obtain directions and advice. YouTube is known for communities for skill sharing and learning \cite{YouTubeParticipatoryCulture}. SHWM videos can provide instructions, guidelines, or advice during the pandemic. \textit{Opportunity for nurturance} is a provision in which the person feels responsible for another's wellbeing. During COVID-19, YouTubers may communicate with their viewers to show their concern for their health and wellbeing. 
\par
The annotation of social provisions in SHWM videos was through crowd-sourcing on Mechanical Turk\footnote{https://www.mturk.com/}. The six social provisions of SHWM videos were rated by participants who watched SHWM videos. The annotation generated six binary variables for each video to represent whether the video provides each of the six social provisions. Logistic regression models were built with the video style and COVID-mentioning as independent variables and the six social provision variables as dependent factors. The predictive models reveal how videos in different styles and mentioning COVID-19 in different degrees affect the video's social provisions. How different video styles mention COVID-19 was also compared by the Wilcoxon test (posthoc uses Dunn's test with Bonferroni adjustment).

\subsection{Viewer Engagement}
Parasocial relationships could help people shield against loneliness \cite{GardnerSocialSnacking, HartmannParasocialInteractionWellbeing, GileParasocialInteraction}. Interacting with the videos can simulate more realistic social connections to the YouTubers \cite{RasmussenParasocialInteractionYouTubeCelebrities}. Video watching, liking, and commenting reflect viewers' participation in the parasocial interactions on YouTube, which may mitigate the effects of loneliness \cite{KhanSocialMediaEngagement, HartmannParasocialInteractionWellbeing}. Interactions such as commenting and liking on other social media are also considered positive indicators of mental wellbeing \cite{ParkFacebook, Verduyn2015PassiveEvidence.}. RQ1 and RQ2 explore what videos were created for SHWM and how they offered social provisions for people in social distancing. RQ3 seeks to capture each social provision's effects on the interactions with SHWM videos and their potential effects on loneliness by measuring viewer engagement metrics. User engagement in online services is defined as ``the quality of the user experience'' that motivates people to interact \cite{LalmasUserEngagement}. Lehmann et al. identified \textit{popularity}, \textit{activity}, and \textit{loyalty} as three key metrics to measure users' engagement \cite{LalmasUserEngagement}. On YouTube, these measurements indicate viewers' participation and behaviors in the parasocial interactions to fulfill the need for social interaction \cite{KhanSocialMediaEngagement}. A video's popularity can be reflected by the number of views, likes, and comments it received. Popularity measurements reflect if providing a social provision allowed the YouTubers to reach and support more viewers. Activity is the average viewer's activeness in the interaction with the video and the YouTuber. The frequencies of likes and comments a video received per 100 views were collected to measure this dimension. Activity measurements indicate viewers' activeness in the parasocial interactions. Loyalty metrics are how often the users return to the social media site. Since it is difficult to collect viewers' watching history from YouTube, this work leveraged user comments' sentiment as an alternative factor of loyalty. NRC Word-Emotion Association Lexicon \cite{MohammadLexicon} was used to count emotional words in viewer comments. Comment emotions evaluate viewers' emotional connection to the YouTuber's content. Although viewer engagement metrics do not directly measure how much loneliness SHWM videos can mitigate, they reflect the degree to which viewers were attracted by the videos and willing to participate in the social interactions on YouTube \cite{KhanSocialMediaEngagement}. Ordinary least squares models (OLS) were built to measure whether videos with different social provisions resulted in different viewer engagement. The model used the six provision variables (dummy variables) to predict popularity, activity, and comment emotion measurements. However, it is widely recognized that the number of followers a creator has is a decisive factor for their content popularity \cite{Chatzopoulou2010AYouTube, UntoldStoryClones}. Viewer engagement is significantly correlated with subscriber count. Therefore the multivariate models included subscriber count to investigate whether the six social provisions added an extra layer of effects on viewer engagement besides creators' popularity.

\section{Data and Analysis Methods}
The SHWM video data was collected with YouTube Data API\footnote{https://developers.google.com/youtube/v3} between Jun 3 and Jun 5, 2020. The data crawling included videos with publishing dates between Mar 11 (the declaration of state emergency) and May 15 (the first round of reopening in most U.S. states). The authors chose to collect the data at least 20 days after the video was published because it left enough time for the newer videos to get views \cite{Chatzopoulou2010AYouTube}. Videos published in countries other than the U.S. were not considered because of different COVID-19 quarantine time-frame and difficulties in analyzing non-English videos. The keyword ``\#StayHome \#WithMe'' was used to retrieve an initial list of videos between the start and end dates. Video metadata includes the title, description, duration, publishing date, view count, like and dislike count, comment count, subscriber count, and the first 200 available comments. The initial crawling returned 6375 videos (including videos of all countries and all languages). After raw metadata processing, videos without channel information, shorter than 5 seconds, or being viewed less than 100 times were eliminated. The sanity check also removed private videos, deleted videos, and videos with broken links. After the cleaning, 4733 videos were excluded from the dataset. The remaining 1642 videos constituted the dataset for data encoding. These videos were from 695 different YouTubers, viewed 206,319,247 times, and attracted 3,301,412 likes and 310,879 comments. The median duration of the videos was 10.73 minutes ($mean=28.70$, $SD=65.28$).

\subsection{Categorizing Video Styles}
\begin{figure}[ht]
\centering
\Description[Examples of videos in nine video styles.]{Examples of nine video styles including artistic, challenge, chatting, game, homelife, how-to, religious, review, and story.}
\scalebox{.75}{
    \begin{tabular}{lllll}
        \begin{minipage}[t]{.55\linewidth}
        \includegraphics[width=\linewidth]{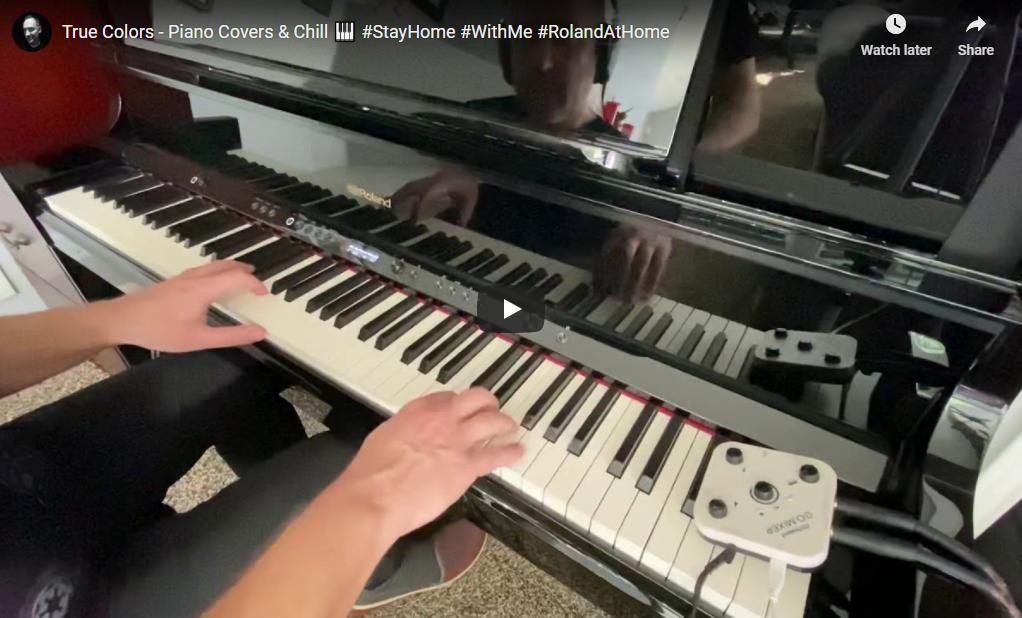}\\
        (a) True Colors - Piano Covers \& Chill \#StayHome \#WithMe \#RolandAtHome
        \end{minipage}
        &
        \begin{minipage}[t]{.55\linewidth}
        \includegraphics[width=\linewidth]{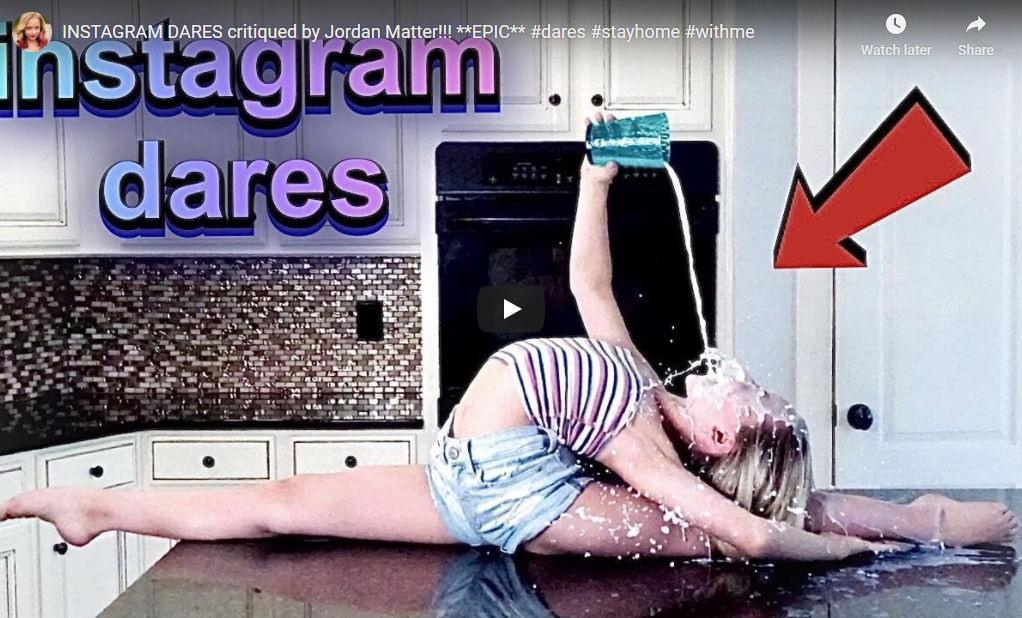}\\
        (b) INSTAGRAM DARES critiqued by Jordan Matter!!! EPIC \#dares \#stayhome \#withme
        \end{minipage}
        \\[9ex]
        \begin{minipage}[t]{.55\linewidth}
        \includegraphics[width=\linewidth]{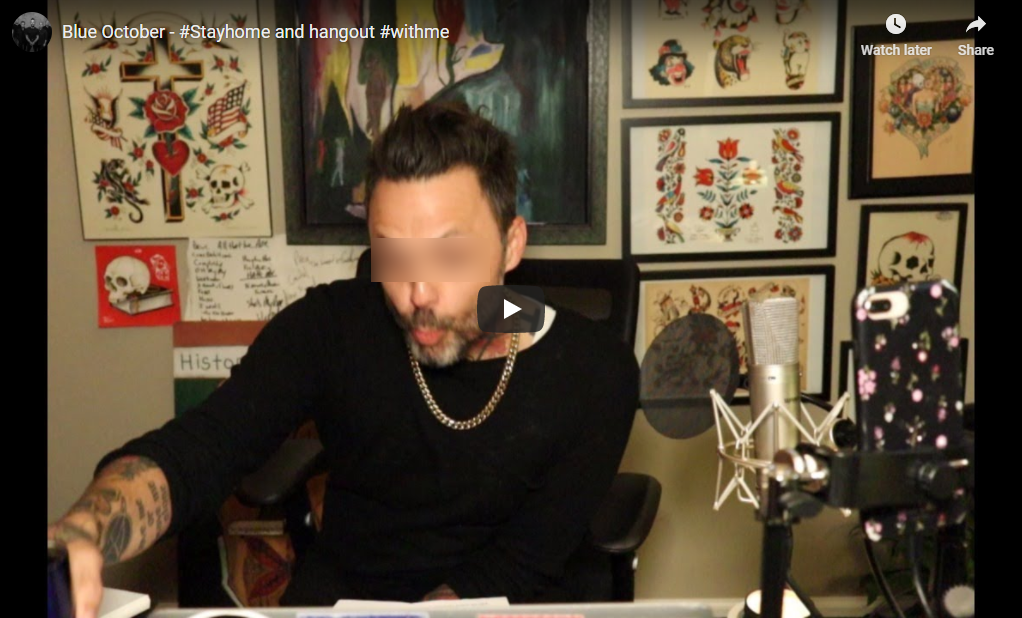}\\
        (c) Blue October - \#Stayhome and hangout \#withme
        \end{minipage}
        &
        \begin{minipage}[t]{.55\linewidth}
        \includegraphics[width=\linewidth]{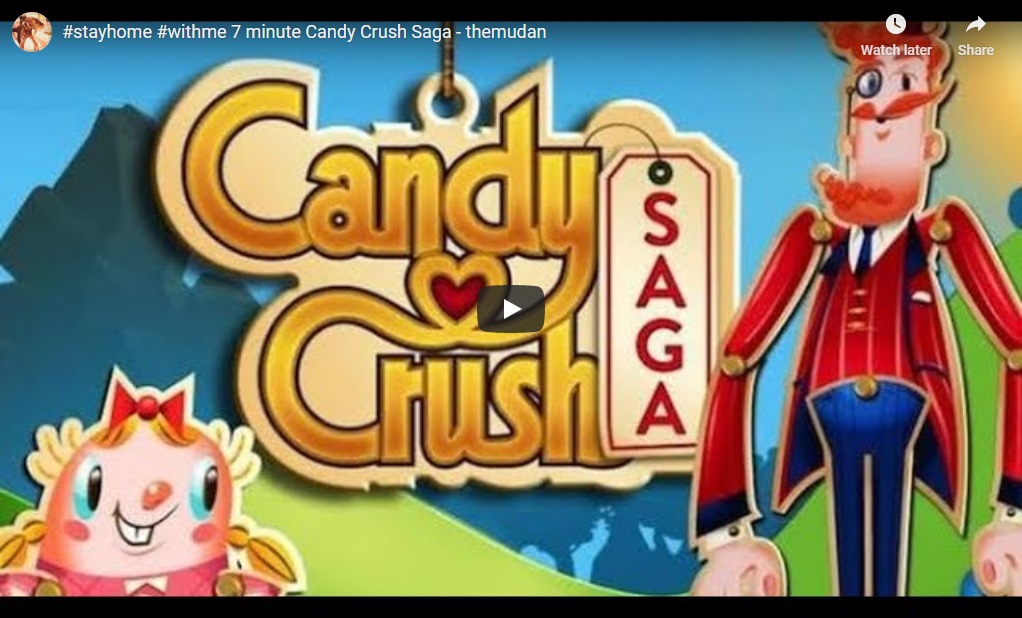}
        (d) \#stayhome \#withme 7 minute Candy Crush Saga - themudan
        \end{minipage}
        \\[6ex]
        \begin{minipage}[t]{.55\linewidth}
        \includegraphics[width=\linewidth]{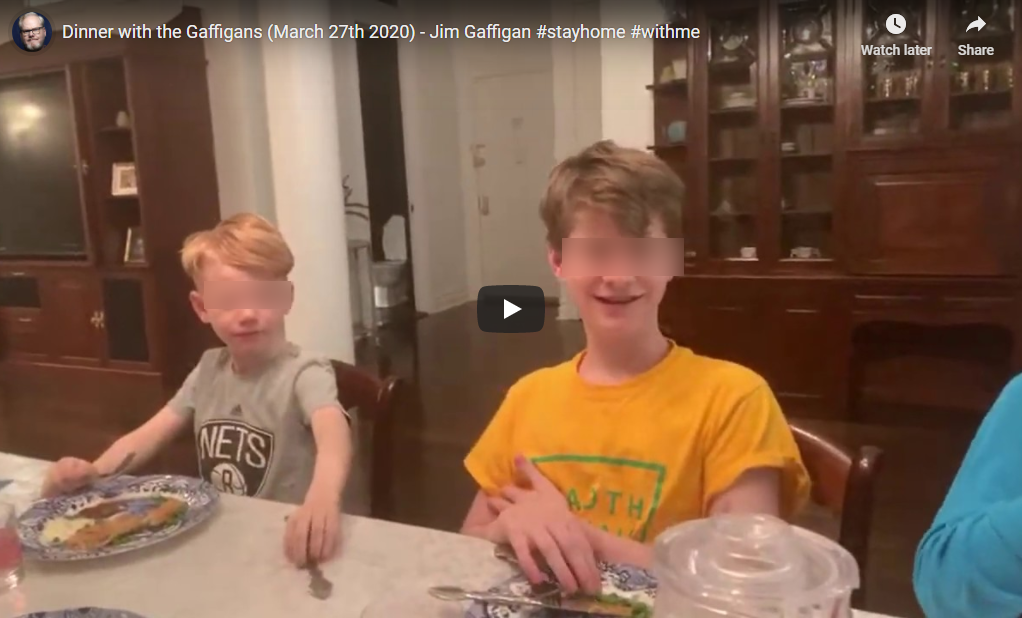}
        (e) Dinner with the Gaffigans March 27th 2020 - Jim Gaffigan \#stayhome \#withme
        \end{minipage}
        &
        \begin{minipage}[t]{.55\linewidth}
        \includegraphics[width=\linewidth]{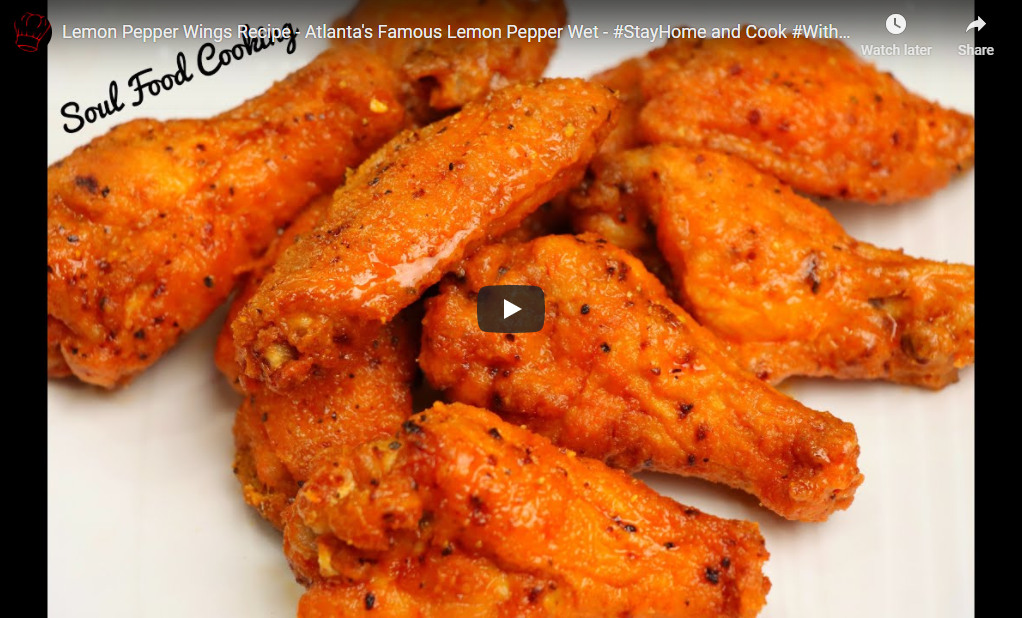}
        (f) Lemon Pepper Wings Recipe - ... - \#StayHome and Cook \#WithMe
        \end{minipage}
        \\[9ex]
        \begin{minipage}[t]{.55\linewidth}
        \includegraphics[width=\linewidth]{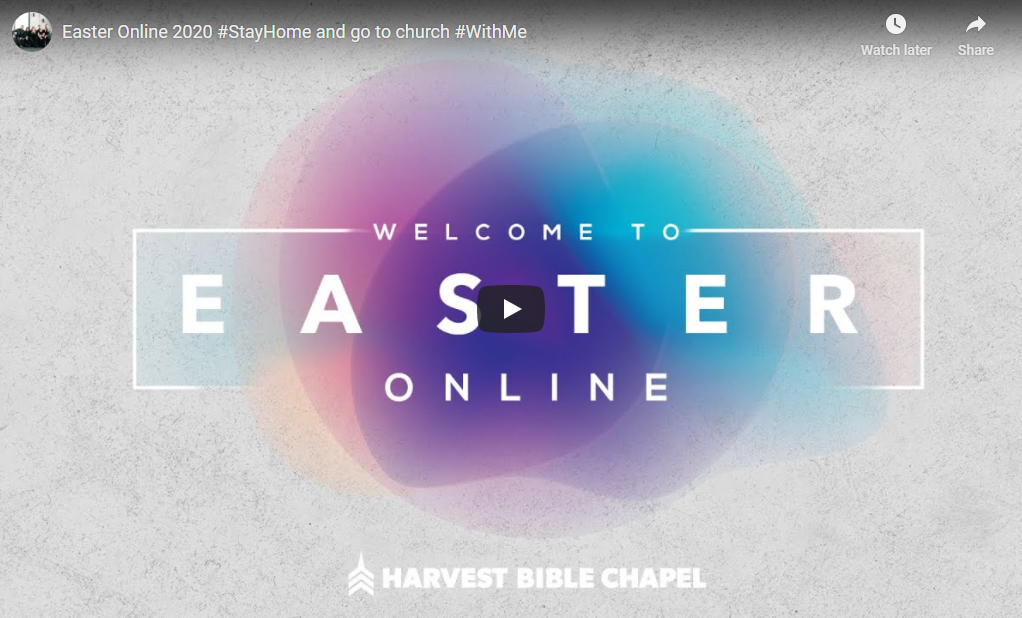}
        (g) Easter Online 2020 \#StayHome and go to church \#WithMe
        \end{minipage}
        &
        \begin{minipage}[t]{.55\linewidth}
        \includegraphics[width=\linewidth]{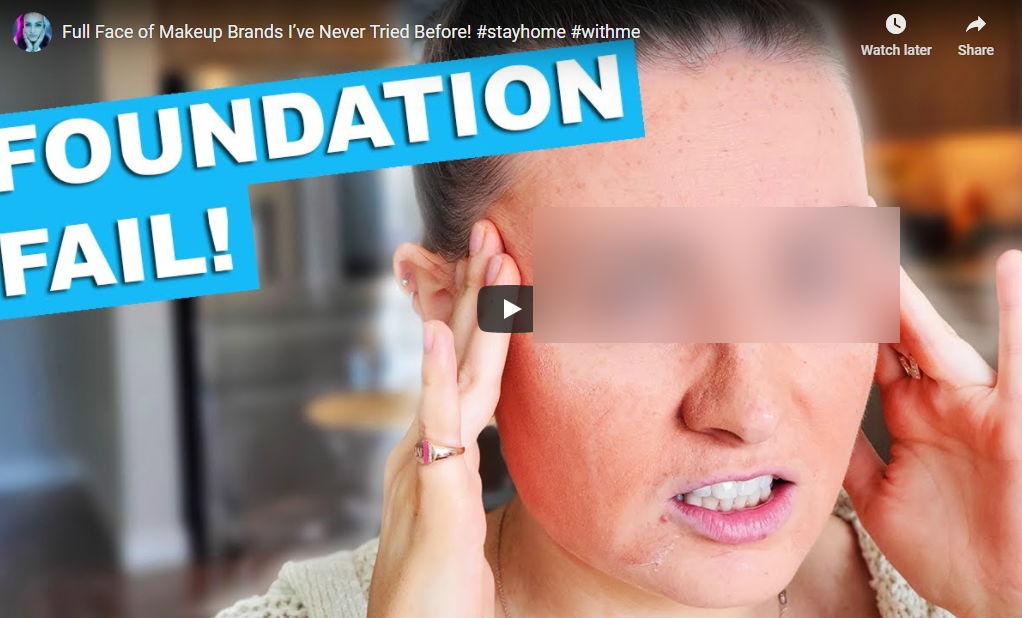}
        (h) Full Face of Makeup Brands I’ve Never Tried Before! \#stayhome \#withme
        \end{minipage}
        \\
        \begin{minipage}[t]{.55\linewidth}
        \includegraphics[width=\linewidth]{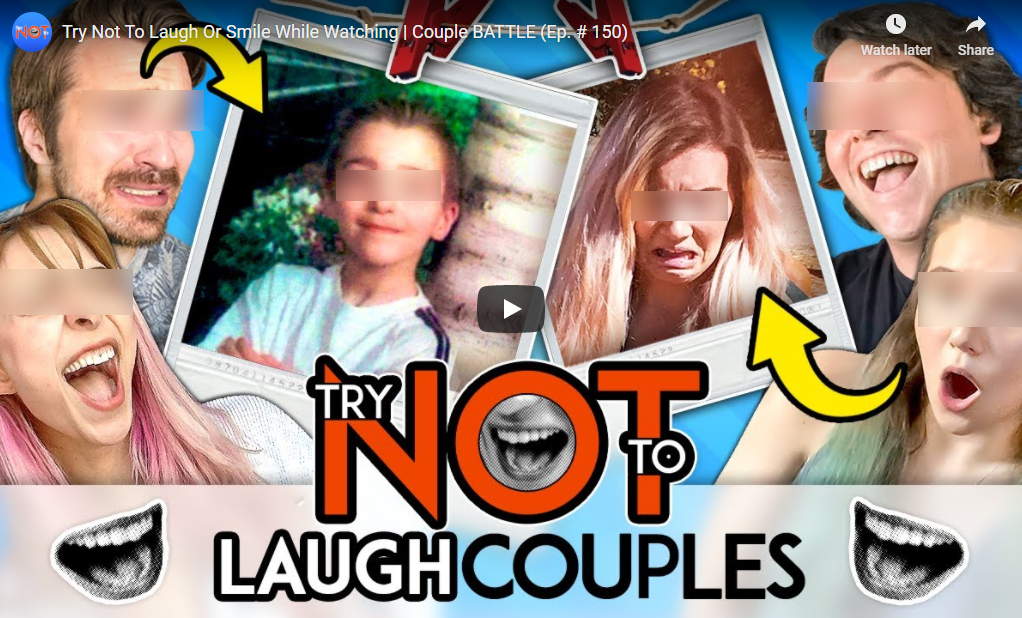}
        (i) Try Not To Laugh Or Smile While Watching Couple BATTLE Ep. \# 150
        \end{minipage}
    \end{tabular}
}
\caption{Example videos in 9 video styles: (a) artistic, (b) challenge, (c) chatting, (d) game, (e) homelife, (f) how-to, (g) religious, (h) review, and (i) story.}
\label{fig:video_examples}
\end{figure}

Grounded theory approach \cite{Charmaz2015GroundedTheory} was applied to identify video styles. In the open coding stage, one author watched 100 randomly selected SHWM videos and generated 100 notes. The author specifically inspected what activities were done and the videos' production style (e.g., livestream, animation, photos/memes). For example, in a video of ``cooking with me'', the author noted ``a video to teach cooking, the creator talked about cooking steps, and the video shared the cooking hobby''. After that, the notes were summarized into emerging video styles through affinity diagramming. Open encoding generated nine video styles. A discriminant analysis was conducted through closed-encoding to validate the video styles' accuracy in representing SHWM videos. Four authors used the nine styles to tag another 300 videos, with each video tagged by two authors. Tagging results were then compared to identify discrepancies. When an inconsistency was identified, revisions to the style definitions were made. After the discussion, the authors finalized a reliable categorization of nine video styles for crowd-sourcing annotation. Researchers reached a substantial agreement on the video styles during the discriminant analysis ($Fleiss' kappa = 0.725$, $p<0.001$). The resulted styles are presented in Table \ref{tab:video_styles} and example videos can be found in Figure \ref{fig:video_examples}.

\begin{table}[ht]
    \caption{Nine video styles identified from the grounded theory analysis.}
    \Description[Nine video styles identified from the grounded theory analysis.]{Description of nine video styles including artistic, challenge, chatting, game, homelife, how-to, religious, review, and story.}
    \centering
    \scalebox{.75}{
        \begin{tabular}{|m{.15\linewidth}|p{\linewidth}|}
            \hline
            Style & Description \\
            \hline
            \textit{artistic} & A video of music, art (drawing/painting/carving/etc.), performance, or animation\\
             \hline
             \textit{challenge} & A video showing exciting activities or participating in a unique challenge to grab attention (e.g. stunts, self-challenges)\\
             \hline
             \textit{chatting} &  A video or livestream checking in with or interacting with the audience (e.g. chatting, reading viewer comments, or doing activities)\\
             \hline
             \textit{game} & A gameplay or a recorded livestream game video\\
             \hline 
             \textit{homelife} &  A video or vlog showing family, homelife, or lifestyle\\
             \hline
             \textit{how-to} & A how-to video or knowledge video that gives step-by-step guidelines to complete a task or explains a specific subject to the audience (e.g. cooking, learn language, safety tips)\\
             \hline
             \textit{religious} & A religious video showing prayers or praying\\
             \hline
             \textit{review} & A video reviewing or recommending products or services\\
             \hline
             \textit{story} &  A video sharing an interesting story, pictures (i.e. photos, memes), or video clip(s)\\
             \hline
        \end{tabular}
    }
    \label{tab:video_styles}
\end{table}

\subsection{Annotating Video Styles, COVID-19 Mentioning, and Social Provisions}
782 participants from MTurk were recruited to complete the annotation task. All participants were from the U.S. and have completed at least 5000 tasks with an accuracy of 97\% or higher. Each participant was asked to watch a video for at least three minutes. The task consists of three questions that annotate video styles, COVID-19 mentioning levels, and whether it provides each of the six social provisions, plus one quality check question to estimate if the participants watched the video carefully. Q1 asked the participants to pick one of the nine video styles (or ``none of the above'') that best describes the video content (Table \ref{tab:video_styles}). Q2 was a quality-control question that asked the participants to select the YouTube video category from three randomly generated options. Q3 asked the participants to pick all applied social provisions, with an extra choice of ``none of the above.'' The provision descriptions were rephrased for easy understanding (see Table \ref{tab:provision_coding} for Q3 options). Participants were asked to rate the video by considering someone who regularly views this kind of video. Q4 was a question to annotate the degree to which the video mentions COVID-19, coronavirus, or pandemic by selecting one of ``none'', ``low'', ``medium'', and ``high''. Participants were asked to rate based on their perception of whether the video discussed COVID-19. No participants tagged more than 1/10 of all 1642 videos. To ensure the validity of the data annotation, the authors filtered out and republished tasks that either not watched enough time as required, with unanswered questions, with conflicting answers (e.g. ``none of above'' and one above option are checked at the same time), or with apparently wrong answers for Q2. The task was performed in two rounds to ensure agreement on the choices. In the first round, each video was tagged by three different participants. A task was considered to reach an agreement if at least two participants picked the same video style or ``none of the above''. If a video has three different style answers, it was tagged by two additional participants. Then the video style was determined by the one that was selected by three participants. Videos without agreement were also identified. The average rating of Q4 was used as the value of COVID-19 mentioning (0 for ``none'' and 3 for ``high''). The variables for six provisions were set to 1 if more than half of the participants picked that provision in Q3; otherwise, they were set to 0. The agreed results of Q1, Q3, and Q4 were saved as variable values to perform multivariate analysis (see Table \ref{tab:variables}).

\begin{table}[ht]
    \caption{Annotations of social provisions in Q3}
    \Description[Annotations of social provisions in Q3.]{The options of six social provisions in Q3. The question asks MTurk participants to select all applicable social provisions.}
    \centering
    \scalebox{0.75}{
        \begin{tabular}{|m{.18\linewidth}|m{.92\linewidth}|}
            \hline
            Social\linebreak Provisions & Provision Coding \\
            \hline
            attachment & The video creator gives a sense of emotional security and closeness to the audience.\\
            \hline 
            integration &  The video creator shares common or specialized hobbies and interests, or shares entertaining content, activities, or experiences with the audience.\\
            \hline
            reassurance & The video creator demonstrates special skills and abilities in hopes to be acknowledged, or acknowledges the audience’s thoughts and comments.\\
            \hline
            alliance & The video creator expresses a willingness to help anytime, i.e. being available to help with the audience’s problems or difficulties.\\
            \hline
            guidance &  The video creator gives step-by-step instructions, guidelines, or advice on a subject that they are knowledgeable in. \\
            \hline
            nurturance & The video creator feels responsible for and interested in the audience's wellbeing.\\
            \hline
        \end{tabular}
    }
    \label{tab:provision_coding}
\end{table}
\begin{table*}[ht]
    \caption{The measured factors and their respective variables used in the data analysis}
    \Description[The measured factors and their respective variables.]{The factors measured in the study. Variables include video style, COVID-19 mentioning, six social provisions, popularity, activity, and comment emotion measurements.}
    \centering
    \scalebox{0.7}{
        \begin{tabular}{|m{.1\textwidth}|p{.2\textwidth}|p{.1\textwidth}|p{.5\textwidth}|}
            \hline
            & Factor & Variable & Definition\\
            \hline
            \multirow{2}{.15\textwidth}{Video \\Creation} & Video style & $style$ & The video style in Table \ref{tab:video_styles} that picked by more than half of the participants in Q1.\\
            \cline{2-4}
            & COVID-19 Mentioning & $cov$ & Average rating of COVID-mentioning in Q4\\
            \hline
            \multirow{6}{.15\textwidth}{Social \\Provisions} & Attachment & $attachment$ & \multirow{6}{.4\textwidth}{Value is set to 1 if more than half of the participants checked the provision in Q3, otherwise 0.} \\
            \cline {2-3}
            & Social Integration & $integration$ &\\
            \cline {2-3}
            & Reassurance of worth & $reassurance$ &\\
            \cline {2-3}
            & Reliable alliance & $alliance$ &\\
            \cline {2-3}
            & Guidance & $guidance$ &\\
            \cline {2-3}
            & Opportunity for nurturance & $nurturance$ & \\
            \hline
            \multirow{3}{.15\textwidth}{Popularity} & View count & $view$ & \multirow{3}{.4\textwidth}{Number of times the video has been watched/liked/commented}\\
            \cline {2-3}
            & Comment count & $comment$ & \\
            \cline {2-3}
            & Like count & $like$ & \\
            \hline
            \multirow{2}{.15\textwidth}{Activity} & Like rate & $like\_rate$ & \multirow{2}{.4\textwidth}{Likes/comments a video received per 100 views} \\
            \cline {2-3}
            & Comment rate & $comment\_rate$ & \\
            \hline
            \multirow{2}{.15\textwidth}{Comment \\Emotion} & Positive emotion & $positive\_score$ & \multirow{2}{.4\textwidth}{The frequency of positive/negative emotional words in the comments} \\
            \cline {2-3}
            & Negative emotion & $negative\_score$ & \\
            \hline
        \end{tabular}
    }
    \label{tab:variables}
\end{table*}

\subsection{Measuring Viewer Engagement}
This work considers three aspects of viewer engagement to probe how SHWM viewers parasocially interact with YouTubers: popularity, activity, and comment emotion \cite{LalmasUserEngagement}. The number of views, likes, and comments constitute the measurements of popularity. View count ($view$), like count ($like$), and comment count ($comment$) are common measurements to assess to what degree the video reached viewers \cite{Chatzopoulou2010AYouTube}. Activity is measured by how many likes and comments a video got for every 100 views -- like rate ($like\_rate$) and comment rate ($comment\_rate$). The like rate is the difference between like and dislike count per 100 video views, calculated using Eq. \ref{eq:1}. The comment rate is the number of comments a video received per 100 video views (Eq. \ref{eq:2}). Those two factors reflect the activeness of parasocial interactions with the YouTubers. To measure comment emotions, emotional words in NRC Word-Emotion Association Lexicon \cite{MohammadLexicon} was used to count how many positive and negative emotional words were used in the viewer comments. Positive (or negative) emotion score of a comment is calculated by the total number of positive (or negative) words in the collected comments divided by the number of counted comments ($positive\_score$ and $negative\_score$, Eq. \ref{eq:3}) \cite{YuWorldCupNRC}. The frequencies of positive and negative words in comments reflect whether viewers express favorable or unfavorable sentiment towards the video, which implies the positivity of parasocial interactions with the YouTuber. 
\par
\begin{equation}
\Description[Calculation of like rate.]{Like rate equals likes minus dislikes divided by view count times 100.}
\label{eq:1}
  like\_rate=\frac{likes-dislikes}{view\_count}*100
\end{equation}

\begin{equation}
\Description[Calculation of comment rate.]{Comment rate equals comment count divided by view count times 100.}
\label{eq:2}
  comment\_rate =\frac{comment\_count}{view\_count}*100
\end{equation}

\begin{equation}
    \label{eq:3} 
    \Description[Calculation of positive and negative comment score.]{Positive score equals number of positive words divided by total counted comments. Negative score equals number of negative words divided by total counted comments.}
    posi(nega)tive\_score =\frac{number\_of\_posi(nega)tive\_words}{total\_counted\_comments} 
\end{equation}

\section{Findings}
The participants spent an average of 5.42 minutes on the task ($SD=2.52$). 14 were annotated as ``none of the above'' per consensus, and 140 videos reached no agreement on the video style. Since the videos that could not be categorized constituted less than 1\%, the authors decided to exclude them from the analysis. For the 140 (8.5\%) no-agreement videos, the authors reviewed a portion and posited that these videos have contents from multiple styles (e.g., presenting fine arts and talking about steps or reviewing products and showing family activities at the same time). Since this study focuses on the social provisions of clearly discernible video styles, the authors decided to exclude the 140 videos to avoid confusion and uncertainty in the prediction models. For all included videos, Fleiss' kappa between the first three crowd workers' style annotations was 0.566 ($p<0.001$). To verify encoding reliability, the authors randomly selected 50 videos annotated by the participants and determined their styles. There was a substantial agreement between the author annotated styles and the crowd-annotated video styles ($Fleiss' kappa = 0.727$, $p<0.001$). The lower agreement between individual MTurk participants suggests that viewers may be interested in different aspects of the same video. For example, a yoga video at home might be annotated as \textit{homelife}, but some participants may be interested in how the YouTubers did it and tagged it as a \textit{how-to}. But annotating the video by the majority rule allowed the authors to decide the closest style to describe the video.

\subsection{RQ1: What are \#StayHome \#WithMe Videos?}
RQ1 probes what SHWM videos were created and how they relate to the pandemic. The analysis consists of an initial comparison with the overall YouTube video category distribution \cite{Che2015ACharacteristics} and a grounded-theory encoding of SHWM videos. The degree to which each video mentions the pandemic is also rated to explore whether SHWM videos were made for sharing COVID-19 information.

\subsubsection{Video Styles of \#StayHome \#WithMe}
\begin{figure}[ht]
    \centering
    \includegraphics[width=.88\linewidth]{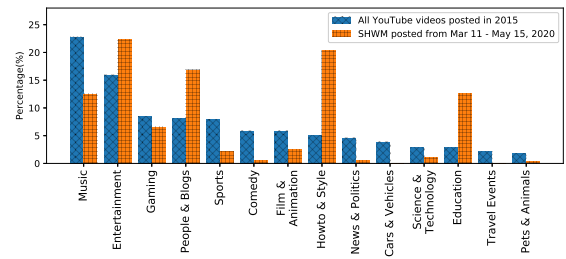}
    \caption{Comparison of category distributions between SHWM videos and 2015 YouTube video distribution (data source: \cite{Che2015ACharacteristics})}
     \Description[Comparison of category distributions between SHWM videos and 2015 YouTube video distribution.]{There are more videos in the categories of Entertainment, People & Blogs, Howto & Style, Education in \#StayHome \#WithMe than the overall distribution of YouTube videos in 2015.}
    \label{fig:youtube_distribution}
\end{figure}

\begin{figure}[ht]
  \centering
     \includegraphics[width=.88\linewidth]{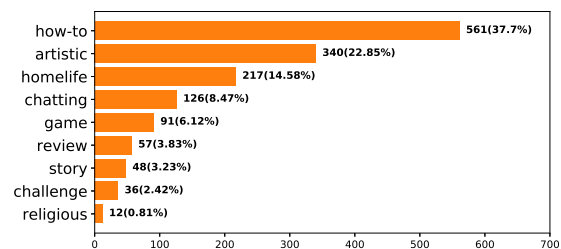}
    \caption{Number of videos in each video styles}
  \label{fig:style_distribution}
\end{figure}

The comparison revealed that in contrast to the distribution in \cite{Che2015ACharacteristics} (overall video category distribution in 2015), SHWM had more videos in ``Entertainment'', ``Howto \& Style'', ``People \& Blogs'', and ``Education'' (Figure \ref{fig:youtube_distribution}). There were 20.36\% ``Howto \& Style'' videos and 12.63\% ``Education'' videos in SHWM, in contrast to their 5.1\% and 2.9\% overall proportions. There were 22.38\% and 16.87\% in the categories of ``Entertainment'' and ``People \& Blogs'', higher than the 16.0\% and 8.1\% in the overall distribution. This result indicates that during the COVID-19 quarantine, YouTubers used \#StayHome \#WithMe hashtags to entertain, share advice and tips with, and present personal life and vlogs to their viewers to help them avoid or mitigate loneliness. The encoding of video styles reveals similar results: \textit{how-to}, \textit{artistic}, \textit{homelife}, \textit{chatting}, and \textit{game} were the five most common video styles (Figure \ref{fig:style_distribution}). The style distribution suggests that the most common ways for YouTubers to build parasocial relationships and help with loneliness were showing step-oriented and knowledge building videos, making entertaining videos by demonstrating artistic or gaming skills, and showing a video of home activities or chatting to the audience (See Table \ref{tab:style_top} for example videos).

\begin{table*}[ht]
    \centering
    \scalebox{1}{
    \begin{tabular}{m{0.2\textwidth}m{0.7\textwidth}}
         \cline{2-2}
         \includegraphics[width=\linewidth]{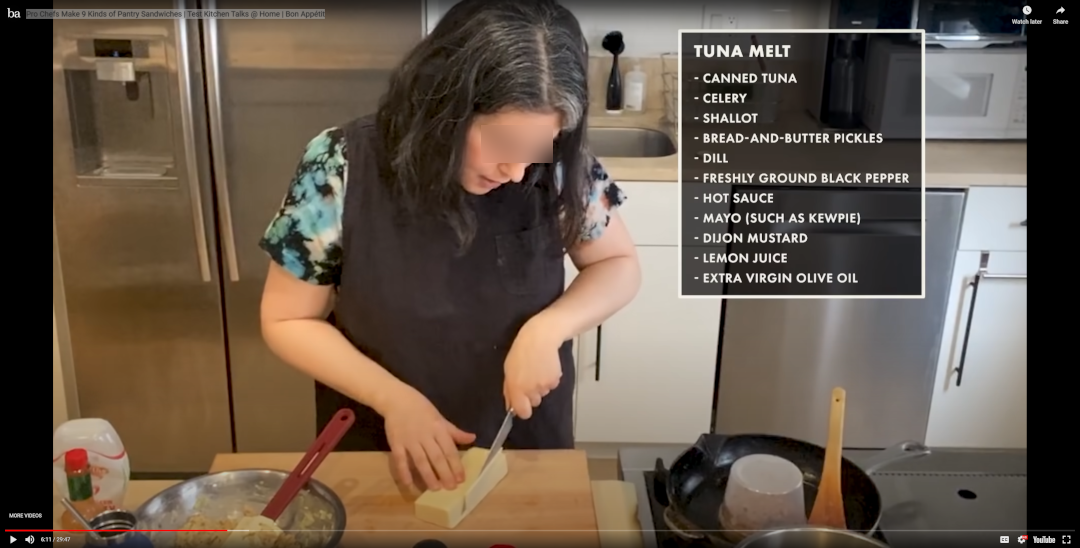} & Title: Pro chefs make 9 kinds of pantry sandwiches | Test Kitchen Talks @ Home | Bon Appétit.
         \newline
         Content: 9 chefs show how to cook pantry sandwiches at home. The YouTubers cook while explaining each step. 
         \newline
         Provisions: guidance and integration
         \\
         \cline{2-2}
         \includegraphics[width=\linewidth]{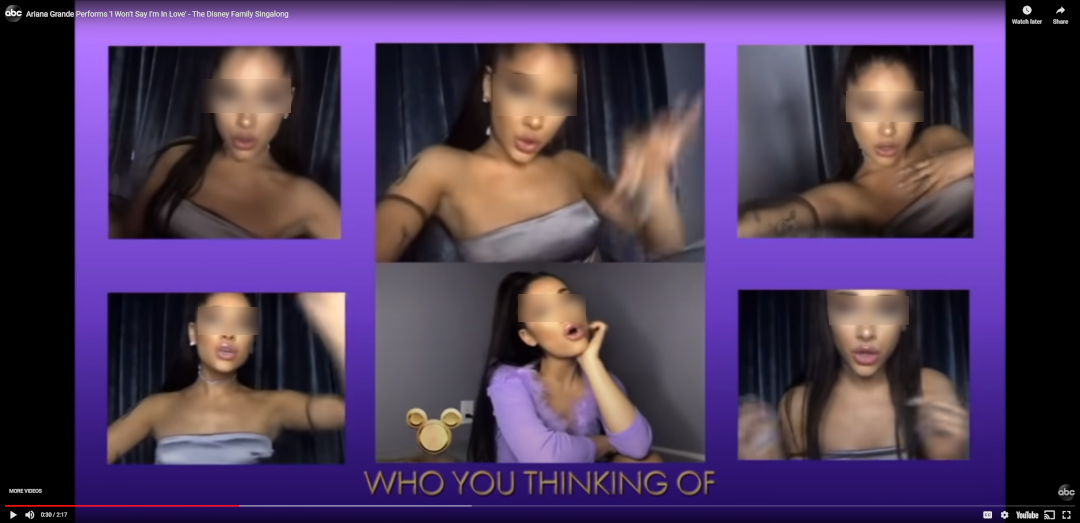} & Title: Ariana Grande Performs 'I Won't Say I'm In Love' - The Disney Family Singalong
         \newline
         Content: A YouTuber sings a Disney song while acting funny postures. The song is about the experience of falling in love.
         \newline
         Provisions: attachment and integration
         \\
         \cline{2-2}
         \includegraphics[width=\linewidth]{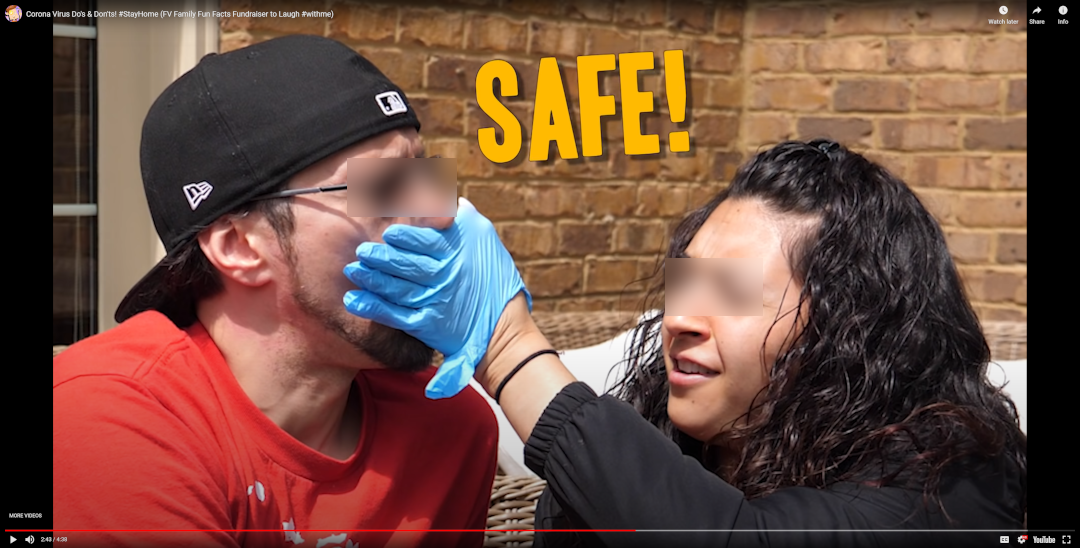} & Title: Corona Virus Do's \& Don'ts! \#StayHome (FV Family Fun Facts Fundraiser to Laugh \#withme)
         \newline
         Content: The YouTubers use exaggerated and funny at-home performance to encourage viewers to wash hands, not to shake hands, and share foods wearing gloves. The YouTubes tell viewers to stay safe. 
         \newline
         Provisions: attachment, guidance, and nurturance
         \\
         \cline{2-2}
         \includegraphics[width=\linewidth]{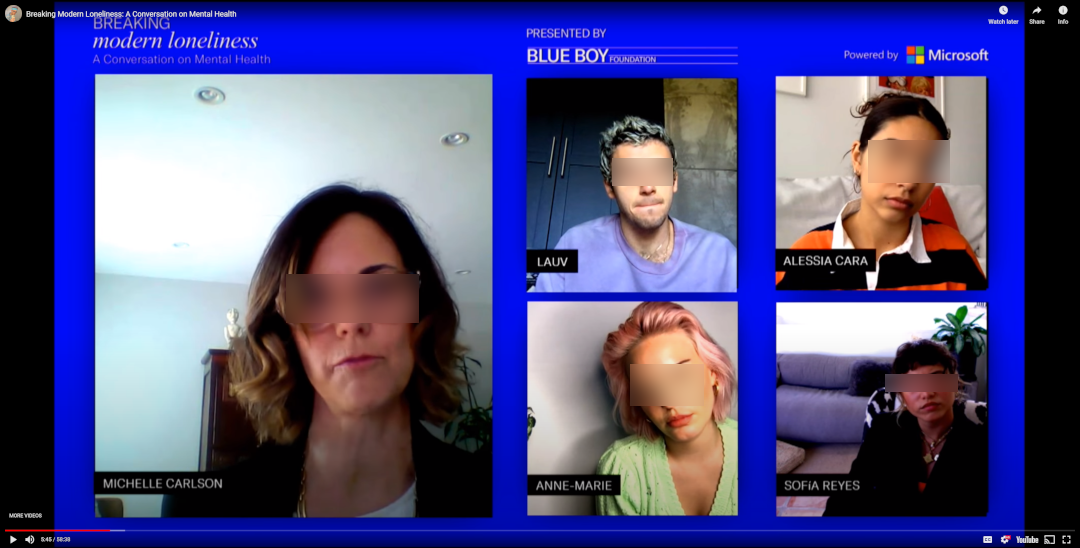} & Title: Breaking Modern Loneliness: A Conversation on Mental Health
         \newline
         Content: Five speakers live a chat to share stories, talk about coping with difficulties, and answer questions. 
         \newline
         Provisions: alliance, attachment, and nurturance\\
         \cline{2-2}
         \includegraphics[width=\linewidth]{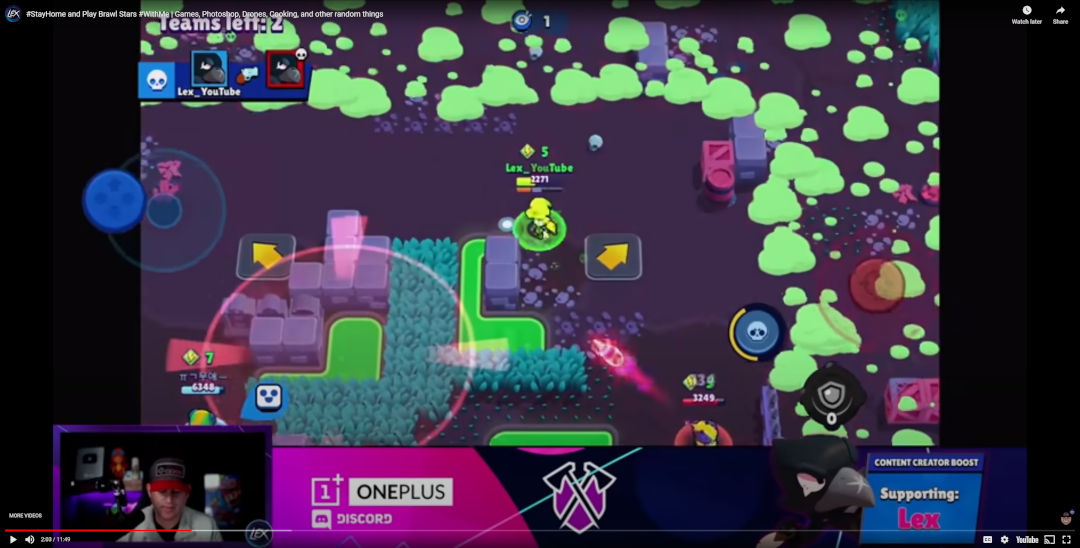} & Title: \#StayHome and Play Brawl Stars \#WithMe | Games, Photoshop, Drones, Cooking, and other random things
         \newline
         Content: A live streaming gamer plays an online multiplayer game. The gamer explains his moves and game progress while showing his excitement to the audience. 
         \newline
         Provisions: attachment and integration\\
         \cline{2-2}
    \end{tabular}
    }
    \caption{Videos with highest like count in the top 5 video styles. From top to bottom: \textit{how-to}, \textit{artistic}, \textit{homelife}, \textit{chatting}, and \textit{game}. The social provisions were annotated by MTurk participants.}
     \Description[Example videos in top five styles.]{Example videos of how-to, artistic, homelife, chatting, and game.}
    \label{tab:style_top}
\end{table*}
\subsubsection{Mentioning the COVID-19 Pandemic}
\begin{figure}[ht]
    \centering
    \includegraphics[width=.7\linewidth]{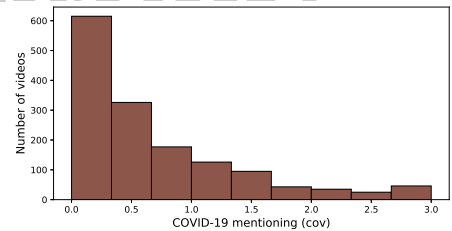}
    \caption{The distribution of COVID-19 mentioning. On X-axis, 0 for ``not mentioned'' and 3 for ``highly mentioned''}
    
     \Description[The distribution of COVID-19 mentioning.]{The COVID-19 mentioning scores of most videos fall between 0 (none) and 1 (low).}
    \label{fig:cov_distribution}
\end{figure}

\#StayHome \#WithMe were two YouTube hashtags popularized during the COVID-19 pandemic. However, the authors noticed that most SHWM did not intensively mention the pandemic. The average rating of COVID-19 mentioning was 0.55, between ``none'' and ``low'' mentioning. Only 21 videos (1.41\%) were rated as highly mentioned COVID-19 pandemic, whereas 615 videos (41.33\%) did not even mention the pandemic (see distribution in Figure \ref{fig:cov_distribution}). Wilcoxon test with posthoc pair-wise comparisons ($\chi^2(8)=141.70$, $p<0.0001$) suggested that \textit{homelife} and \textit{chatting} had significantly higher $cov$ than \textit{artistic}, \textit{how-to}, \textit{challenge}, \textit{story} and \textit{game} videos (all $p<0.0064$). This result suggests that YouTubers did not discuss or spread pandemic-related information when they sought to mitigate viewers' loneliness during \#StayHome \#WithMe. Instead, SHWM videos had more content for skill-learning and entertaining to redirect viewers' attention from COVID-19. For creators who mentioned COVID-19 more, their videos showed homelife or chat with the audience. SHWM videos didn't intensively mention COVID because YouTubers felt discussing this stressor in the video would negatively impact viewers' mental states and wellbeing \cite{EllisSocialStress}.

\subsection{RQ2: Video Styles, COVID-19 Mentioning, and Social Provisions}
\begin{figure}[ht]
  \centering
      \includegraphics[width=.88\linewidth]{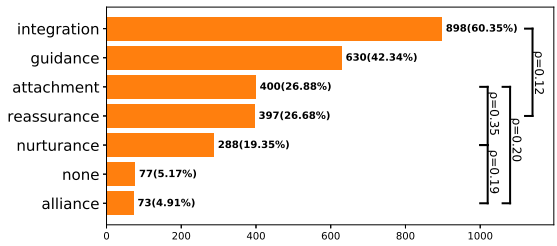}
    \caption{Number of videos providing each of the social provisions. Vertical bars show correlations.}
    \label{fig:provision_distribution}
\end{figure}

\begin{figure}[ht]
    \includegraphics[width=.32\linewidth, angle=270]{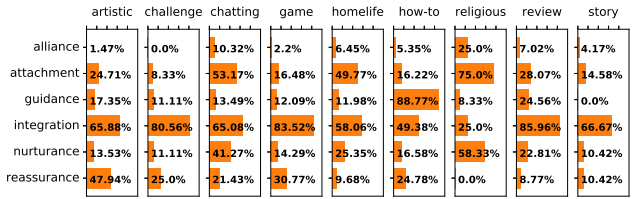}
    \caption{The percentages of videos providing the six social provisions by different video styles}
     \Description[The percentages of videos providing the six social provisions.]{The figure shows the percentage of videos in each of the nine video styles that provide each of the six social provisions. }
    \label{fig:style_provision}
\end{figure}

\begin{figure}[ht]
    \includegraphics[width=.95\linewidth]{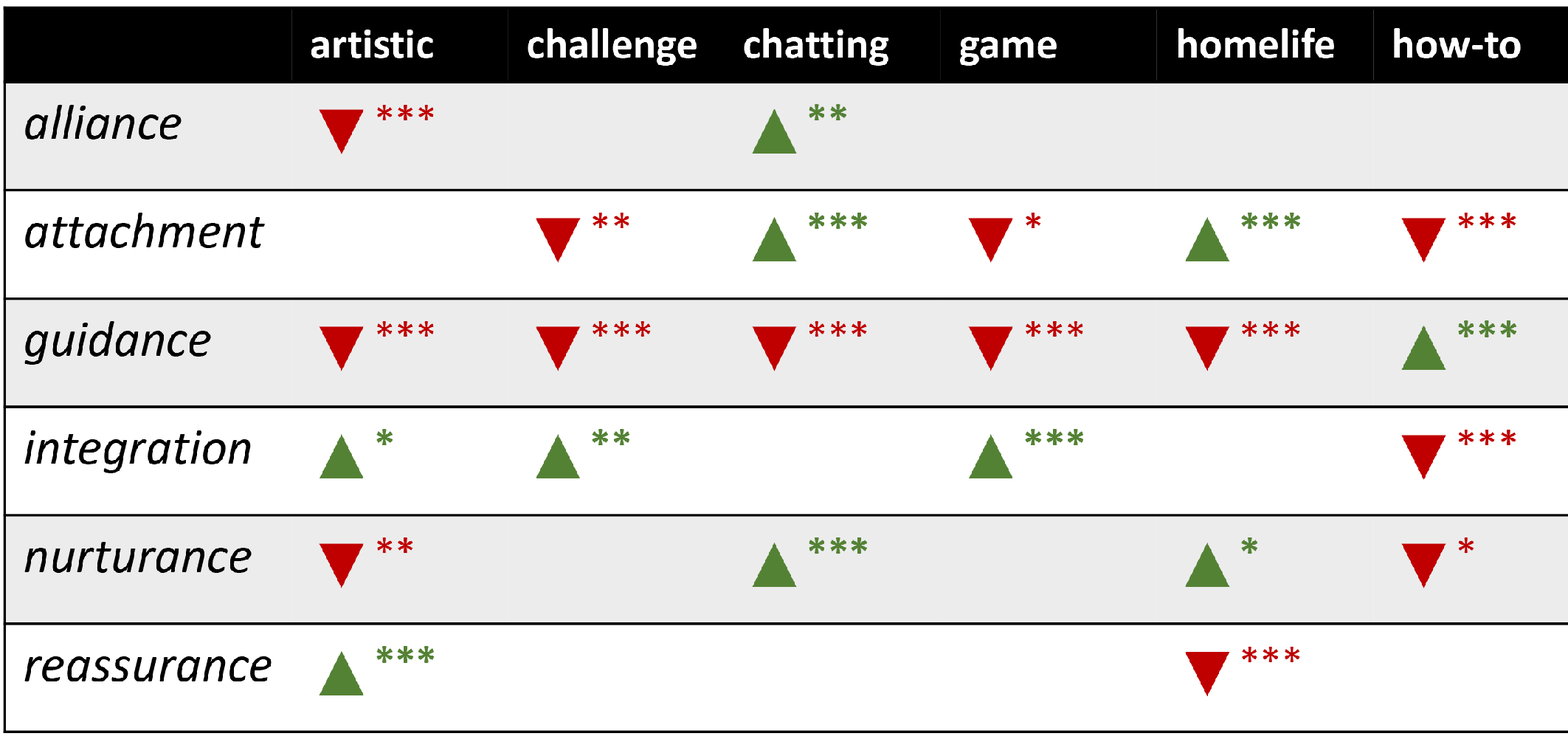}
    \caption{Significant associations between video styles and social provisions. Green triangles are positive associations. Red ones are negative associations. p* < 0.05, p** < 0.01, p*** < 0.001.}
     \Description[Significant associations between video styles and social provisions.]{Alliance is associated with chatting and religious. Attachment is chatting, homelife, and religious. Guidance is associated with how-to. Integration is associated with artistic, challenge, game, and review. Nurturance is associated with chatting, homelife, and religious. Reassurance is associated with artistic.}
    \label{fig:style_association_provision}
\end{figure}

RQ2 probes how SHWM videos associate with the social provisions \cite{Cutrona1987TheStress}. Social integration and guidance were the most-supported provisions (Figure \ref{fig:provision_distribution}). Attachment and reassurance of worth had a similar amount of videos. Opportunity for nurturance and reliable alliance, the two social provisions come from family-like relationships, were supported by the fewest. Among 1488 videos, 77 were tagged not to associate with any social provision. Spearman's $\rho$ test suggested weak positive correlations between $attachment$ and $nurturance$ ($\rho=0.35$, $p<0.0001$), $attachment$ and $alliance$ ($\rho=0.20$, $p<0.0001$), $alliance$ and $nurturance$ ($\rho=0.19$, $p<0.0001$), and $reassurance$ and $integration$ ($\rho=0.122$, $p<0.0001$). Nominal logistic regression (LR) was used to predict each of the six social provisions by $style$ and $cov$. The alpha value to decide model significance was 0.0083 (0.05/6, after Bonferroni correction). One-sided Fisher's exact test was the posthoc to identify associations between styles and provisions ($\alpha=0.0056$ after Bonferroni correction). Figure \ref{fig:style_provision} shows the percentages of videos in different video styles offering each social provision. Figure \ref{fig:style_association_provision} illustrates significant association between video styles and social provisions.

\subsubsection{Social integration}
Social integration describes a friend-based network in which people share common interests and concerns. In SHWM, social integration was a pervasive social provision provided by most video styles. Except for \textit{religious} videos, the participants rated more than 49\% videos in all other styles to provide social integration (Figure \ref{fig:style_provision}). The LR model revealed a collective significant effect of $style$ and $cov$ on $integration$ ($\chi^2(9)=98.45$, $R^2=0.05$, $p<0.0001$), with $style$ and $cov$ having significant variable effects ($p_{style}<0.0001$, $p_{cov}=0.0017$). Posthoc revealed that \textit{artistic}, \textit{challenge}, \textit{game}, and \textit{review} had significantly more videos providing social integration (Figure \ref{fig:style_association_provision}). While \textit{how-to} and \textit{religious} had significantly fewer videos related to social integration than other styles. It was also noticed that videos with high COVID-19 mentioning were less associated with $integration$ ($coe=-0.25$, $\chi^2=9.82$). These results indicate that sharing interests was the most common way to support loneliness during SHWM. The interests covered a wide range of topics: from artistic content to showing stunts and self-challenges, from live gaming to recommending products or services. YouTubers made videos of interests and hobbies that entertain viewers and mitigate loneliness. 

\subsubsection{Guidance} 
Guidance refers to a mentor-like relationship in which people obtain advice and assistance from a trustworthy person. YouTube communities include professional creators who can impart skills and knowledge through step-by-step instructions or knowledge explanation.  \cite{Burgess2018YouTube:Culture, YouTubeParticipatoryCulture}. The LR model revealed that the $style$ and $cov$ had a significant positive effect on $guidance$ ($\chi^2(9)=903.18$, $R^2=0.45$, $p<0.0001$, $p_{style}<0.0001$, $p_{cov}=0.0290$). In contrast to social integration, guidance was a provision primarily supported by \textit{how-to} videos (Figure \ref{fig:style_association_provision}). Posthoc showed that only the \textit{how-to} style had a significantly higher proportion (88.77\%) of videos that provide guidance. While guidance videos were created during COVID-19, they were not particularly mentioning this disaster ($coe=-0.26$, $\chi^2=4.77$, $p_{cov}=0.0290$). This result indicates that YouTubers shared general skills and knowledge to give viewers guidance and a mentor-like relationship, but the knowledge was not about health and safety information.

\subsubsection{Attachment}
Attachment describes family-like relationships that create a sense of safety and closeness to dispel emotional loneliness. LR model suggested a collective significant effect between $style$, $cov$, and $attachment$ ($\chi^2(9)=192.20$, $R^2=0.11$, $p<0.0001$, $p_{style}<0.0001$, $p_{cov}<0.0001$). Posthoc suggested that \textit{chatting}, \textit{homelife}, and \textit{religious} videos had significantly higher proportions of videos providing attachment, while \textit{challenge}, \textit{game}, and \textit{how-to} had significantly fewer (Figure \ref{fig:style_association_provision}). In \textit{chatting} videos, YouTubers interacted with the audience through face-to-face talking, which gave the audience a sense of attachment. \textit{Homelife} videos were at-home activities to show YouTuber's intimacy. Religious videos provided attachment by showing prayers or quotes from religious scriptures. In contrast to social integration and guidance, videos that offered attachment appeared to be more related to COVID-19. The coefficient for $cov$ was 0.52 ($\chi^2=35.8$, $p_{cov}<0.0001$). These results suggest that attachment in SHWM videos was provided by talking to the audience, showing YouTubers' home activities, or giving a religious prayer. YouTubers concerned about the audience's wellbeing during the pandemic were more likely to offer the attachment to help them mitigate loneliness.
\par
\subsubsection{Opportunity for nurturance and reliable alliance}
Similar to attachment, the opportunity for nurturance and reliable alliance are derived from family-based relationships. The former describes a relationship where one feels responsible for the other’s wellbeing. The latter emphasizes one can count on the other under any circumstances. However, these two provisions were supported by the fewest SHWM videos. Since they had similar LR and posthoc results, this section explains these two provisions together. LR model suggested that $style$ and $cov$ had an significant effect on $nurturance$ ($\chi^2(9)=149.89$, $R^2=0.10$, $p<0.0001$, $p_{style}<0.0001$, $p_{cov}<0.0001$) and $alliance$ ($\chi^2(9)=48.25$, $R^2=0.08$, $p<0.0001$, $p_{style}<0.0001$, $p_{cov}<0.0001$). Posthoc analysis suggested that \textit{homelife}, \textit{chatting}, and \textit{religious} videos had significantly more videos with $nurturance$ (Figure \ref{fig:style_association_provision}). The coefficient of $cov$ in the model was 0.81 ($\chi^2=74.01$, $p_{cov}<0.0001$), which indicates that higher COVID-19 mentioning was associated with a higher chance of providing nurturance. For $alliance$, posthoc analysis showed that \textit{chatting} and \textit{religious} videos were associated with $alliance$ (Figure \ref{fig:style_association_provision}). The coefficient of $cov$ in the LR model was 0.58 ($\chi^2=15.7$, $p<0.0001$), indicating the reliable alliance was also correlated with higher COVID-19 mentioning. These results indicate that similar to attachment, the opportunity for nurturance and reliable alliance were provided by videos of chatting and religious content. \textit{Homelife} videos also provided opportunity for nurturance. The concerns over COVID-19 encouraged YouTubers to make videos to provide these two family-like provisions, despite they were only provided by fewer SHWM videos.

\subsubsection{Reassurance of Worth} 
Reassurance of worth is a relationship in which one's skills and abilities are acknowledged by others, which is usually achieved by offering help or understanding to others \cite{Cutrona1987TheStress}. The LR analysis suggested that $style$ was a significant impact factor to this provision, but $cov$ did not affect $reassurance$ ($\chi^2(9)=139.61$, $R^2=0.08$, $p<0.0001$, $p_{style}<0.0001$, $p_{cov}=0.8621$). Posthoc showed that \textit{artistic} is the only video style positively associated with this social provision. 163 out of 340 \textit{artistic} videos were tagged to support reassurance of worth, while \textit{homelife}, \textit{religious}, \textit{review}, and \textit{story} videos had significantly fewer videos with this provision (Figure \ref{fig:style_association_provision}). Considering that artistic videos also provided social integration, participants considered that YouTubers presented artistic content or performance to entertain viewers while hoping their skills and abilities can be acknowledged.

\subsection{RQ3: Social Provisions and Viewer Engagement}
RQ3 explores how videos with different social provisions affected viewer engagement to imply how different types of parasocial relationships affect viewers' interactions with the video and their effects on mitigating loneliness. For each social provision, the authors compared if demonstrating a social provision increased or decreased the three popularity measurements, two activity measurements, and two comment emotion measurements. The comparison was performed within each of the nine video style groups. Ordinary Least Squares (OLS) model was used as the multivariate analysis method. To ensure the model's comprehensiveness, the model also incorporated subscriber count ($sub$) as an additional independent factor. Therefore, social provisions (dummy variables) and subscriber count (numerical) were the independent variables to predict the dependent variables of viewer engagement. 125 videos were excluded from the viewer engagement analysis because their creator disabled comment and/or like functions. The alpha to determine model significance was 0.0071 (0.05/7, after Bonferroni correction). The correlation between $sub$ and each provision was tested with Spearman's $\rho$ test to avoid multicollinearity. No significant correlation was detected.

\subsubsection{Popularity}
In the OLS model which predicted video popularity, $view$ and $like$ did not have any association with any social provision variables (only correlated with $sub$). But for \textit{how-to}, \textit{homelife}, and \textit{review} videos, besides $sub$, one or more social provisions had significant effects on the comment count ($comment$). For \textit{how-to} videos, $comment$ was significantly associated with $attachment$ ($F(7)=9.09$, $R^2=0.11$, $p<0.0001$). OLS model showed that the coefficient of attachment was 56.01 ($t_{attachment}=2.08$, $p_{attachment}=0.0384$) in predicting how-to videos' comment count, which suggested a positive association between attachment and comments. For \textit{homelife} videos, opportunity for nurturance also had a positive coefficient of 73.45 ($F(7)=25.08$, $R^2=0.50$, $p<0.0001$, $t_{nurturance}=2.39$, $p_{nurturance}=0.0179$) in predicting $comment$. For \textit{review} videos, the model suggested that reassurance of worth had a coefficient of 139.95 ($F(7)=7.11$, $R^2=0.53$, $p<0.0001$, $t_{reassurance}=2.26$, $p_{reassurance}=0.0287$) in predicting comments. While social provisions did not affect the view and like amount of SHWM videos, videos in \textit{how-to}, \textit{homelife}, and \textit{review} styles attracted more comments by providing attachment, nurturance, or reassurance. It implies that despite social provisions didn't help SHWM videos to reach more audience; they had a positive effect on encouraging more viewers to leave a comment. It is also interesting to note that for \textit{how-to} and \textit{review} videos, the social provisions that positively affected commenting were family-like provisions. However, they were not the main social provisions offered by SHWM. 

\subsubsection{Activity}
Like rate ($like\_rate$) and comment rate ($comment\\\_rate$) were the two metrics to measure viewers' activeness in the interactions with SHWM videos. OLS model suggested a collective significant effects of social provisions and $sub$ on $like\_rate$ and $comment\_rate$ in \textit{how-to} and \textit{artistic} groups. For \textit{how-to} videos, there was a significant effect of social provisions on $like\_rate$ ($F(7)=3.18$, $R^2=0.04$, $p=0.0027$). The model suggested that providing alliance made \textit{how-to} videos attracted more likes per 100 views ($coe=1.05$, $t_{alliance}=2.01$, $p_{alliance}=0.0444$). The variable of $nurturance$ had significant positive effects on $comment\_rate$ of \textit{artistic} videos ($F(7)=4.17$, $R^2=0.09$, $p=0.0002$). The coefficient of $nurturance$ in the model was 1.46 ($t_{nurturance}=2.25$, $p_{nurturance}=0.0253$). These results indicated that although alliance and nurturance were the least provided social provisions in SHWM, offering those two provisions helped \textit{how-to} videos to gain more likes and \textit{artistic} videos to gain more comments. The result implies that showing alliance in \textit{how-to} videos and nurturance in \textit{artistic} videos encouraged viewers to more actively participate in the parasocial interactions on YouTube. 

\subsubsection{Comment Emotion}
\begin{figure}[ht]
    \includegraphics[width=.95\linewidth]{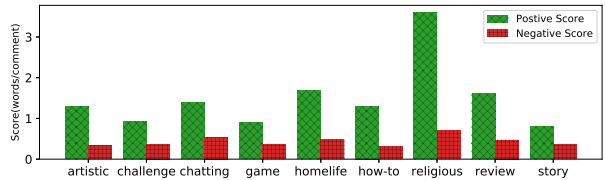}
    \caption{The frequencies of positive and negative emotional words in videos in different styles}
    \label{fig:emotion_distribution}
\end{figure}

The comment analysis predicts the frequencies of emotional word in viewer comments by the factors of social provisions and the subscriber count. 70245 comments from 1135 videos (at most 200 for each video) were included in this analysis (353 videos had no comment or disabled commenting). OLS model suggested no social provisions had significant effects on $positive\_score$ or $negative\_score$ in any of the nine style groups. Therefore the authors cannot conclude that providing different social provisions had an effect on comment emotions. The authors then chose to examine whether different video styles and COVID-19 mentioning had an effect on comment emotions. The OLS model indicated significant effects of $style$ and $cov$ on $positive\_score$ and $negative\_score$. $Cov$ showed a significant effect in the model which predicts $negative\_score$ ($coe=0.05$, $t=2.22$, $p=0.0267$). $Style$ had significant effects on $positive\_score$ and $negative\_score$ ($p_{positive\_score}<0.0001$, $p_{negative\_score}=0.0004$). Pairwise comparison with Dunn method suggested \textit{religious} and \textit{homelife} videos had significantly higher positive word frequencies than \textit{game}, \textit{challenge}, \textit{artistic}, \textit{chatting}, \textit{how-to}, and \textit{story} videos (all $p<0.0089$, Figure \ref{fig:emotion_distribution}). Providing specific social provisions didn't alter viewers' emotional expression in the comments. But viewers' comments in videos which mentioned COVID-19 more were more negative, which suggest that COVID-19 content may evoke negative feelings. In contrast, viewers' comments to \textit{religious} and \textit{homelife} videos, the two styles associated with attachment and nurturance, were more positive than other video styles. As SHWM is a movement for promoting positive mental attitudes and avoiding loneliness, these results indicate that videos in these two styles led to more positive viewer reactions and emotions to the YouTubers. 

\section{Discussion}
\subsection{\#StayHome \#WithMe Did Not Escalate Pandemic Attention}
The analysis of SHWM suggested YouTubers sought to de-escalate disaster attention in SHWM videos. Social media is used as a tool to produce, spread, and access disaster information during emergencies and disasters. Prior research focused on examining the affordances of social media, including YouTube \cite{OwensDisasterYouTube, BaschYouTubeEbola}, in disseminating and receiving disaster information and serving as an emergency management tool \cite{HoustonSocialMediaDisaster, LindsaySocialMediaDisasters, PalenInformatingCrisis}. For example, Twitter enhanced public situational awareness \cite{PalenTwitterDiaster}; Facebook was a source of community and official information \cite{BirdFloodingFacebook, ChauhanWildfireFacebook}; and Reddit was used for risk perception and speculation \cite{YunanRedditCrisis}. However, more exposure to the overwhelming COVID-19 news via social media caused negative mental states such as stress and depression \cite{GroarkeLonelinessPandemic, EllisSocialStress}. Prior works on social media and disaster mental health mainly examined media's roles in mental health surveillance \cite{KarmegamReviewMentalHealthSocialMedia} offering mental health service \cite{HoustonSocialMediaDisaster}, and enhancing community ties \cite{ProcopioCrisisSupportGroup}.
\par
The trending of \#StayHome \#WithMe implies an alternative affordance of YouTube that contrasts prior usage of social media in disasters. Instead of accentuating COVID-19 information and raising situational awareness, this work found that SHWM YouTubers utilized parasocial relationships to create a virtual space to allow people to reduce the pandemic stress. In contrast to the information source on Twitter and Facebook who provide immediate disaster information, creators of \#StayHome \#WithMe did not prioritize information on the pandemic. They provided various social provisions to redirect or de-escalate viewers' attention from disaster-related stressors. The average COVID-19 mentioning was between ``none'' and ``low''. Only 1.41\% of the videos were rated as highly related to the pandemic, whereas 41.33\% videos did not even mention the pandemic in videos. Videos with relatively higher COVID-19 mentioning were not about alarming the disaster or directing disaster management; instead, they sought to provide supportive and comforting provisions such as attachment, alliance, and nurturance. The most prominent social provisions in SHWM -- social integration and guidance -- were negatively associated with COVID-19, indicating YouTubers demonstrated hobbies or taught skills as an approach to de-escalate pandemic attention and prioritize connection. 
\par
The prolonged social distancing and overwhelming COVID information put people, especially young adults, at risk for increased loneliness and depression \cite{EllisSocialStress, GroarkeLonelinessPandemic}. SHWM movement indicates new design opportunities for using video sharing to mitigate the loneliness and stress caused by disaster information. Besides informing emergencies during a disaster, it is also imperative to call on social media platforms and services to consider providing content that is not disaster-intensive to allow people to experience normality and reduce negative mental states. The findings of this work suggest that SHWM YouTubers would not put the disaster situation at the center of the parasocial relationships with viewers. Instead, they tended to share entertaining, educational, and comforting content. This affordance of video-sharing platforms is valuable during a crisis like COVID-19 when people cannot socialize and do everyday things they usually enjoyed. The non-disaster-focused videos that supplement absent social provisions can support viewers who need to relax and reduce loneliness and stress. This characteristic of YouTube and video-sharing communities should be considered and leveraged in future technology and service designs for loneliness relief and mental health support during disasters.

\subsection{From Parasocial Relationships to Social Provisions in a Disaster}
The analysis of social provisions in SHWM videos implied the manners in which parasocial relationships deliver social connections. The participatory nature of YouTube makes it a platform for ordinary people to contribute content and engage viewers to establish parasocial relationships \cite{Burgess2018YouTube:Culture, WohnParasocialInteraction}. Prior understanding of parasocial relationships in video sharing surrounded how they are formed \cite{ChenDigitalSelfYouTube, KurtinDevelopmentofParasocilIntearctionYouTube} and their effects on the viewers' daily activities \cite{MunnukkaVlogendorsement, SokolovaWhyShouldIBuy, WohnParasocialInteraction}. Although studies showed that parasocial relationships were a source of alternative companionship and can help people shield against loneliness \cite{GileParasocialInteraction, HartmannParasocialInteractionWellbeing, GardnerSocialSnacking}, there is limited knowledge of how YouTube-based parasocial relationships are embodied during social distancing to help with loneliness. 
\par
The examination of SHWM videos under Weiss's typology of provisions revealed the construct of parasocial relationships in SHWM videos and the roles YouTubers played in social distancing. The encoding of SHWM videos suggested that YouTubers sought to offer social connections typically obtained from friend-like, mentor-like, and family-like relationships. The authors found that social integration was a dominant provision in most SHWM videos (provided by 63.35\% videos). Acting as a friend-like character and sharing common interests was the most common parasocial relationship in SHWM. This echoes prior findings that interest sharing helped social media users overcome loneliness and depression \cite{FratamicoRedditLonely, OLearyChatMentalHealth}. Guidance was the second most common provisions available in SHWM videos. 42.32\% of videos provided the guidance provision, most of which were how-to videos. This finding suggested that building informal mentorship through how-to videos \cite{YouTubeParticipatoryCulture} was also a common YouTubers' style to supplement companionship. Family-like provisions, including attachment, nurturance, and alliance, were supported by the fewest videos. These videos shared intimate content, such as at-home activities, live-chatting, and religious prayers, to build intimate provisions and family-like relationships for people who need intimacy and emotional support. Social media is known for connecting families and friends \cite{HoustonSocialMediaDisaster} and sharing emotional content during disasters \cite{KarmegamReviewMentalHealthSocialMedia}. YouTubers act as different roles through their relationships with viewers to offer mental support during COVID-19. SHWM suggested that YouTubers can mitigate disaster loneliness by sharing entertaining content to provide social integration, teaching skills to provide guidance, and showing at-home activities or chatting to provide attachment. 
\par
The correlations between video styles and social provisions suggest practical ways to utilize parasocial relationships to offer loneliness support during a disaster. Viewers can find YouTubers who imitate friend-, mentor-, and family-like connections to mitigate loneliness resulted from social distancing. Social integration was the most prevalent social provision in SHWM videos; therefore, designers may leverage YouTubers' videos to design video applications and services for people who were isolated from friends. People who miss mentor-like relationships, especially children and youth whose schools are closed during the pandemic, can leverage the repository of how-to videos to learn various skills and knowledge. This affordance of YouTube can supplement the inadequate guidance provision. Videos that offer social integration and guidance can also redirect viewers' attention from the crisis. For people who need intimacy and closeness from a family-like relationship, such as people separated from families due to the disasters, video-sharing applications and services may recommend YouTubers' content in which they share their homelife activities or chat with the audience to engender attachment and nurturance. 

\subsection{Encouraging Family-Like Social Provisions in Loneliness-Supporting Videos}
The viewer engagement analysis suggested videos with family-like provisions had better effects on inducing viewers' social interaction participation. Prior studies suggested parasocial interactions can engender intimacy and attachment \cite{ChenDigitalSelfYouTube}, and help people fulfill the social interaction need \cite{KhanSocialMediaEngagement} and mitigate loneliness \cite{HartmannParasocialInteractionWellbeing}. Video views, likes, and comments are metrics of viewer engagement to evaluate parasocial interactions in SHWM \cite{KhanSocialMediaEngagement, RasmussenParasocialInteractionYouTubeCelebrities}. The authors analyzed how social provisions affect viewer engagement to explore social provisions' effects on parasocial relationships. The results suggested that when YouTubers seek to provide parasocial relationships to support mental wellbeing, offering family-like provisions have a better overall effect on increasing viewer interactions and mitigating loneliness. 
\par
There was no evidence that social provisions had effects on helping SHWM videos to reach more viewers. Viewers' emotional expression in comments was also unaffected by the expression of social provisions. However, family-like social provisions demonstrated an overall positive effect on viewers' activeness in parasocial relationships. Attachment increased the number of comments of \textit{how-to} videos, and opportunity for nurturance increased comments of \textit{homelife} videos. For viewers' activity, alliance helped \textit{how-to} videos to receive more likes per 100 views. Providing nurturance allowed \textit{artistic} videos to gain more comments for every 100 views. Although social provisions did not significantly affect comment emotions, the analysis on video styles revealed that \textit{religious} and \textit{homelife} videos -- the two styles bound to attachment and nurturance -- had more positive comments than others. Viewers were more positive after watching videos in those two styles. Prior studies suggested parasocial interactions with YouTubers can avoid and mitigate loneliness \cite{RasmussenParasocialInteractionYouTubeCelebrities, GardnerSocialSnacking, HartmannParasocialInteractionWellbeing}. The authors' findings suggested that SHWM videos that supply attachment, nurturance, and alliance had higher overall viewer engagement. As a result, videos with those provisions positively affected viewers' activeness in the social interactions on YouTube, indicating more potentials to mitigate loneliness. However, these three family-like provisions were also among the least provided social provisions in SHWM videos. Only around 27\% SHWM videos provided attachment and opportunity for nurturance, and only 4.9\% videos provided reliable alliance.
\par
Experiencing close family and friend relationships was helpful to reduce loneliness during COVID-19 \cite{EllisSocialStress}. YouTubers should consider providing more family-like provisions -- by showing intimacy, caring for viewers' wellbeing, and showing a willingness to help -- to increase viewers' interactions and positivity in the parasocial relationships. Platform designers may consider increasing engagement on YouTube for mitigating loneliness by encouraging videos to express family-like provisions. Promising design solutions to grow YouTuber-viewer intimacy include recommending video styles that enhance family-like feelings and implementing communication methods that allow YouTubers to show intimacy and support. Many YouTubers already established family-like profiles among their fans, such as popular homelife vloggers, family-friendly streamers, and many ASMRtists \cite{AndersenShiveriesASMR}. Platforms can invite and encourage these YouTubers to make loneliness-supporting videos during disasters to support public mental wellbeing.

\section{Conclusion and Future Work}
Social media plays an increasingly important role in supporting people's mental wellbeing during a difficult time like COVID-19. Video-sharing platforms like YouTube are conducive for providing social connectedness and reducing loneliness during social distancing. This work examines the \#StayHome \#WithMe movement as a space for YouTubers to help mitigate COVID-19 loneliness. The authors identified video styles and obtained a panoramic understanding of YouTubers' creation activities and parasocial relationships in SHWM. Grounded on Weiss's theory of loneliness, six social provisions were rated by MTurk participants to analyze how videos with different styles and COVID-19 mentioning affect the provisions YouTubers sought to offer. The authors also explored how different social provisions affect video popularity, viewers' activities, and comment emotion, as indicators of participation in parasocial interactions. From the results, the \#StayHome \#WithMe hashtags were primarily used as a space for teaching skills and knowledge, presenting entertaining videos such as artistic presentation and gameplay, and showing homelife activities or chatting with the audience. The videos were less related to the on-going pandemic. The analysis revealed how parasocial relationships supplement social provisions during COVID-19. Social integration was a dominant provision provided by most of the video styles. \textit{How-to} videos supported the need for guidance. \textit{Homelife}, \textit{chatting}, and \textit{religious} videos offered a sense of attachment and nurturance. The viewer engagement analysis suggested that family-like provisions were the least offered provisions, but they positively affected viewer engagement and parasocial interactions. Based on these findings, the authors suggest that SHWM videos sought to de-escalate the mental tension caused by COVID-19. YouTubers offered friend-like and mentor-like provisions the most, while the family-like provisions are supported the least. YouTubers and platform designers should encourage content that offers attachment, nurturance, and alliance during the pandemic to increase parasocial interactions and avoid or mitigate loneliness.
\par
Moving forward, future studies will extend the findings of the present work to advance the knowledge of supporting disaster mental health through video sharing. As user-generated videos will play an increasing role, new studies and designs are needed to understand the interplay between video sharing and mental wellbeing. Follow-up research will extend the findings of this study and develop new design knowledge. For example, one unanswered question in this work is to what degree parasocial interactions can psychologically supplement various social needs during a disaster. The authors do not argue that parasocial relationships with YouTubers can or should replace realistic social interactions with families and friends. However, YouTubers offered an alternative but growingly popular way to let people stay socially connected during disasters; therefore, it requires a more in-depth investigation. Social interactions are easily affected by disasters. YouTube provides an option to satisfy social-emotional needs through parasocial interactions. This work's findings offer a seminal idea regarding the use of YouTube and YouTubers' roles in supporting disaster mental wellbeing. It is necessary to examine YouTube viewers' cognitive and behavioral changes after interacting with YouTube videos during and after disasters. Future work will also investigate new trending video styles such as ASMR and live-streams in obtaining social provisions. These efforts will identify new possibilities of applications and services to utilize video sharing to support mental wellbeing in intensive situations. It is essential to explore their options in intervening in mental health issues of the vulnerable populations.

\bibliographystyle{ACM-Reference-Format}
\bibliography{main.bib}

\end{document}